\documentclass[review]{elsarticle}

\usepackage{graphicx}
\usepackage{dcolumn}
\usepackage{bm}
\usepackage{braket}
\usepackage{graphicx}
\usepackage{epstopdf}
\usepackage{mwe}    
\usepackage{subfig}
\usepackage[T1]{fontenc}
\usepackage{caption} 
\usepackage{amssymb}
\usepackage{amsmath}
\usepackage{booktabs}
\usepackage{tabularx}
\usepackage{multirow}
\usepackage{float}

\newcommand{\tensr}[1]{\bm{\mathsf{#1}}} 

\begin{document}

\begin{frontmatter}
\title{Cascaded Lattice Boltzmann Method based on Central Moments for Axisymmetric Thermal Flows Including Swirling Effects
}

\author{Farzaneh  Hajabdollahi}
\ead{farzaneh.hajabdollahiouderji@ucdenver.edu}

\author{Kannan N. Premnath}
\ead{kannan.premnath@ucdenver.edu}

\author{Samuel W. J. Welch}
\ead{Sam.Welch@ucdenver.edu}

\address{Department of Mechanical Engineering, University of Colorado Denver, 1200 Larimer street, Colorado, 80217 , U.S.A}

\date{\today}

\begin{abstract}
A cascaded lattice Boltzmann (LB) approach based on central moments and multiple relaxation times to simulate thermal convective flows, which are driven by buoyancy forces and/or swirling effects, in the cylindrical coordinate system with axial symmetry is presented. In this regard, the dynamics of the axial and radial momentum components along with the pressure are represented by means of the 2D Navier-Stokes equations with geometric mass and momentum source terms in the pseudo Cartesian form, while the evolutions of the azimuthal momentum and the temperature field are each modeled  by an advection-diffusion type equation with  appropriate local source terms. Based on these, cascaded LB schemes involving three distribution functions are formulated to solve for the fluid motion in the meridian plane using a D2Q9 lattice, and to solve for the azimuthal momentum and the temperature field each using a D2Q5 lattice. The geometric mass and momentum source terms for the flow fields and the energy source term for the temperature field are included using a new symmetric operator splitting technique, via pre-collision and  post-collision source steps around the cascaded collision step for each distribution function. These result in a particularly simple and compact formulation to directly represent the effect of various geometric source terms consistently in terms of changes in the appropriate  zeroth and first order moments. Simulations of several complex buoyancy-driven thermal flows and including rotational effects in cylindrical geometries using the new axisymmetric cascaded LB schemes show good agreement with prior benchmark results for the structures of the velocity and thermal fields as well as the heat transfer rates given in terms of the Nusselt numbers.

\end{abstract}


\end{frontmatter}


\section{\label{1}Introduction}
Fluid motion in cylindrical coordinates with axial symmetry that is driven by rotational effects and/or thermal buoyancy effects arise widely in a number of engineering applications and geophysical contexts~(e.g., \cite{greenspan1968theory,koschmieder1993benard,shevchuk2016modelling,fujimura1997time,lemembre1998laminar}). Some examples of technological applications encountering heat and mass transfer effects in axisymmetric flows include pipeline systems, heat exchangers, solar energy conversion devices, crystal growth and material processing systems, electronic cooling equipment and turbomachinery. Computational methods play an important role for both fundamental studies of the fluid mechanics and heat transfer aspects and as predictive tools for engineering design of such systems. In general, fluid motion in cylindrical coordinates due to swirling effects and buoyancy forces, and accompanied by thermal and mass transport is three-dimensional (3D) in nature. Computational effort for such problems can be significantly reduced if axial symmetry, which arise in various contexts, can be exploited; in such cases the system of equations can be reduced to set of quasi-two-dimensional (2D) problems in the meridian plane. Traditionally, numerical schemes based on finite difference, finite volume or finite elements were constructed to solve the axisymmetric Navier-Stokes (NS) equations for the fluid flow along with the advection-diffusion equation for the energy transport~(e.g., \cite{barbosa2005finite,guardone2007finite}).\par
On the other hand, lattice Boltzmann (LB) methods, which arise as minimal kinetic models of the Boltzmann equation, has attracted much attention and application to a wide range of fluid flows and heat and mass transfer problems ~\cite{he1997theory,succi2001lattice,aidun2010lattice,kruger2017lattice}. They are based on the streaming of the distribution functions of particle populations along discrete lattice directions followed by collision on discrete nodes represented as a relaxation process. The hydrodynamics then arises from the averaged effect of the stream-and-collide steps, where the flow variables are related to the various kinetic moments of the distribution of the particle populations. Hence, the LB methods are characterized as mesoscopic computational  approaches, which have certain unique features and advantages. These include the following. The streaming step is linear and exact and all nonlinearity is modeled locally in the collision step; by contrast, the convective term in the NS equation is nonlinear and nonlocal. As a result, the pressure field is obtained locally in the LB methods, circumventing the need for the solution of the time consuming elliptic Poisson equation as in traditional  methods. The exact-advection in the streaming step combined with the collision step based on a relaxation model leads to a second order accurate method with relatively low numerical dissipation. The kinetic model for the collision step can be tailored to introduce additional physics as necessary and its additional degree of freedom can be tuned to improve numerical stability. Various boundary conditions for complex geometries can be represented using relatively simple rules for the particle populations. Finally, the locality of the method makes it amenable for almost ideal implementation on parallel computers for large scale flow simulations.\par
Following an approach for the solution of the Boltzmann equation in cylindrical coordinates~\cite{mieussens2000discrete}, during the last two decades, various LB schemes for athermal flows (i.e., without heat transfer effects) have been introduced~\cite{halliday2001lattice,premnath2005lattice,lee2006axisymmetric,reis2007modified,zhou2008axisymmetric,chen2008lattice,guo2009theory,huang2009theoretical,li2010improved}. These approaches can be categorized according to the following: (i) Coordinate transformation method~\cite{halliday2001lattice,premnath2005lattice,lee2006axisymmetric,reis2007modified,zhou2008axisymmetric}, in which the axisymmetric mass and momentum equations are reformulated as quasi-2D flow equations in the Cartesian forms with additional geometric source terms and then solved using a LB scheme. (ii) Vorticity-stream function approach~\cite{chen2008lattice}, where LB models are introduced to simulate flows in cylindrical coordinates written in terms of the vorticity and stream function equations. (iii) Radius-weighted formulation~\cite{guo2009theory}, in which a simplified LB method is derived from a discretization of the continuous Boltzmann equation in cylindrical coordinates recast in a radius-weighted form. An analysis of these axisymmetric LB models were performed by~\cite{huang2009theoretical}. Generally, these approaches solve for the axial and radial velocity components in the meridian plane using a popular single relaxation time (SRT) model for the representation of the collision step in the LB scheme. The azimuthal velocity field to represent axisymmetric swirling or rotational flows is computed using an additional LB scheme based on a separate distribution function in a double distribution function (DDF) framework~\cite{guo2009theory,li2010improved}.\par
Further progress in the LB methods for the simulation of axisymmetric thermal flows have been reported in various studies~\cite{peng2003numerical,huang2007hybrid,chen2009simulation,li2009lattice,zheng2010lattice,zheng2010kinetic,li2013multiple,wang2016lattice,wang2014fractional,wang2017modified}. Earlier LB models in this regard~\cite{peng2003numerical,huang2007hybrid} used an hybrid approach, in which the energy equation was solved via a finite difference scheme. Later,~\cite{chen2009simulation,li2009lattice} solved the axisymmetric equation for the temperature field  written in terms of a pseudo-2D advection-diffusion equation with a source term using a LB scheme based on a separation distribution function from that for the flow field. On the other hand,~\cite{zheng2010lattice,zheng2010kinetic} extended the radius-weighted formulation approach for axisymmetric fluid flow~\cite{guo2009theory} for the simulation of thermal energy transport. A fractional-step based LB flux solver for axisymmetric thermal flow was presented in~\cite{wang2016lattice}. All these approaches were based on the common SRT model~\cite{qian1992lattice}, in which, during the collision step, the distribution functions relax to their local equilibria using a single relaxation parameter. This was further extended by the introduction of two tunable parameters as coefficients to the additional gradient terms in the equilibrium distribution functions~\cite{wang2017modified}. Generally, SRT based LB schemes are known to be susceptible to numerical instabilities for convection-dominated flows or fluids with relatively low values of transport coefficients. In order to address this issue, the collision step based on a multiple relaxation time (MRT) model~\cite{d2002multiple} has been constructed, in which raw moments of different orders relax at different rates. Few MRT LB schemes for axisymmetric thermal convective flows have recently been developed~\cite{li2009lattice,li2013multiple,wang2016lattice}.\par
On the other hand, further improvements to the collision step enhancing the flow and thermal transport modeling capabilities can be achieved via the introduction of the cascaded collision model~\cite{geier2006cascaded}. In this approach, the effects of collisions are represented in terms of relaxation of different orders of central moments, which are obtained by shifting the particle velocity, by the local fluid velocity at different rates. As the collision model is prescribed based on a local moving frame of reference, the relaxation steps for successive higher order moments exhibit a cascaded structure. The cascaded collision formulation was shown to be equivalent to considering relaxation to a generalized equilibrium in the rest or lattice frame of reference~\cite{asinari2008generalized}, and was augmented with forcing terms in 2D and 3D in~\cite{premnath2009incorporating,premnath2011three}. Improvements in the numerical properties achieved using such advanced cascaded collision models based on central moments were recently demonstrated~\cite{geier2015cumulant}. A modified formulation based on central moments involving relaxation to discrete equilibria rather than continuous Maxwellian equilibria was also proposed~\cite{de2017non}. In order to accelerate convergence of steady flows, a preconditioned cascaded LB method was constructed and studied in~\cite{hajabdollahi2017improving}, and whose Galilean invariance properties were significantly improved via corrections to equilibria in~\cite{hajabdollahi2018galilean}. The cascaded LB scheme has recently been extended for simulating flows with heat transfer in 2D~\cite{sharma2017new,fei2018modeling} and in 3D in our recent work~\cite{hajabdollahi2018central}. However, for axisymmetric thermal convective flows including rotational effects, no such advanced LB schemes are available in the literature.\par
In this work, we present a new cascaded LB formulation for thermal flows in cylindrical coordinate with axial symmetry, and including rotational effects. The mass, momentum (i.e., axial, radial and azimuthal components) and energy equations rewritten in pseudo-2D Cartesian forms in the meridian plane contain additional geometric source terms, which are included in the respective cascaded LB schemes via a novel symmetric time-split formulation~\cite{hajabdollahi2018Symmetrized}. In this approach, three separate distribution functions are considered:  one for the density, axial and radial momentum components, another one for the azimuthal momentum component and finally third one for the temperature field. Each of the three distribution functions evolves according to a cascaded LB scheme. For this triple distribution functions framework, a two-dimensional, nine velocity (D2Q9) model is used to solve for the axisymmetric NS equations for the axial and radial momentum components, while a two-dimensional five velocity (D2Q5) model is employed to compute the azimuthal momentum and the temperature field, both of whose evolution are represented by advection-diffusion equations with source terms. The use of symmetric operation split formulations based on pre-collision and post-collision source steps for incorporating the geometric source terms for axisymmetric thermal flows including swirl effects leads to a particulary simplified formulation, which is consistent with the classical Strang splitting. The application of central moments based cascaded LB schemes using MRT can enhanced numerical stability of the LB simulations, and the use of symmetric operator splitting yields a scheme that is second order in time, as demonstrated numerically~\cite{hajabdollahi2018Symmetrized}. Such an axisymmetric cascaded LB approach for the simulation of thermally stratified and/or rotating flows in cylindrical geometries can lead to reduced computational and memory costs when compared to a 3D cascaded LB formulation. Several numerical axisymmetric benchmark problems focusing on buoyancy-driven flows and rotational effects are considered to validate our operator-split axisymmetric cascaded LB schemes for thermal flows. These include the Taylor-Couette flow, natural convection in an annulus between two co-axial vertical cylinders, Rayleigh-Benard convection in a vertical cylinder, cylindrical lid-driven cavity flow, mixed convection in a tall vertical annulus and melt flow during Czochralski crystal growth in a vertical rotating cylinder.\par
This paper is organized as follows. In the next section (Sec.~2), cascaded LB methods for axisymmetric thermal flows with swirl effects using a symmetric operator split formulation for the various geometric sources and forces are presented. Numerical results for various benchmark problems are presented and discussed in Sec.3. Finally, the paper concludes with a summary in Sec.4.
\section{Cascaded LB  Methods for Axisymmetric Thermal Convective Flows with Swirling Effects: Symmetric Operator Splitting Formulation} \label{Three}
 We will now present cascaded LB methods based on central moments and MRT for the computation of thermal convective flows in the cylindrical coordinates with axial symmetry, by also taking into account azimuthal rotational/swirling effects. A triple distribution functions based LB approach is considered, where the geometric source terms arising in the pseudo-2D macroscopic equations are represented using symmetric operator splitting around the cascaded collision steps~\cite{hajabdollahi2018Symmetrized}. The solution of the resulting cascaded LB models then yields the local fluid flow variables such as the radial axial and azimuthal velocity fields, pressure (or density) field, and the temperature field in the meridian plane. First, we summarize the macroscopic governing equations for axisymmetric thermal flows subjected to rotation/swirl.
 \subsection{Governing equations for thermal flows in cylindrical coordinates with axial symmetry}
 For incompressible, axisymmetric thermal flows subjected to rotational/swirling effects, the macroscopic governing equations in the cylindrical coordinate system $(r,\theta,z)$ can be written as ~\cite{guo2009theory,zheng2010lattice}
 \begin{subequations}
\begin{eqnarray}
&\partial_t \rho+\partial_r(\rho u_r)+\partial_z(\rho u_z)=-\rho\dfrac{u_r}{r},\label{eq:1a}\\
&\partial_t({\rho u_r})+\partial_r({\rho u_r^2})+\partial_z({\rho u_ru_z})=-\partial_r p+\partial_r(2\mu \partial_ru_r)+\partial_z[\mu (\partial_zu_r+\partial_ru_z)]\\ \nonumber
&+\rho\dfrac{u_\theta^2}{r}-\rho\dfrac{u_r^2}{r}+2\mu\dfrac{\partial_ru_r}{r}-2\mu\dfrac{u_r}{r^2}+F_r^b,\label{eq:1b}\\
&\partial_t({\rho u_z})+\partial_r({\rho u_ru_z})+\partial_z({\rho u^2_z})=-\partial_z p+\partial_r[\mu (\partial_ru_z+\partial_zu_r)]+\partial_z(2\mu \partial_zu_z)\\ \nonumber
&-\rho\dfrac{u_ru_z}{r}+\mu\dfrac{(\partial_zu_r+\partial_ru_z)}{r}+F_z^b,\label{eq:1c}\\
&\partial_t(\rho u_\theta)+\partial_r(\rho u_r  u_\theta)+\partial_z(\rho u_z u_\theta)=\nu\left[\dfrac{\partial^2}{\partial r^2}(\rho u_\theta)+\dfrac{\partial^2}{\partial z^2}(\rho u_\theta)\right]-2\rho\dfrac{u_ru_{\theta}}{r}\\ \nonumber
&+\rho\dfrac{\nu}{r}\partial_ru_\theta-\rho \dfrac{\nu u_{\theta}}{r^2},\label{eq:1d}\\
&\partial_t \phi+\partial_r(u_r\phi)+\partial_z(u_z\phi)=\partial_r(D_\phi \partial_r \phi)+\partial_z(D_\phi \partial_z \phi)-\dfrac{u_r\phi}{r}+\dfrac{D_\phi}{r}\partial_r \phi.\label{eq:1e}
\end{eqnarray}
\end{subequations}
Here, $r$, $z$ and $\theta$ represent the coordinates in the radial, axial and azimuthal directions, respectively; accordingly, $u_r$, $u_z$ and $u_\theta$ denote the fluid velocity components in the respective directions, and $F_r^b$ and $F_z^b$ are radial and axial components of the external body forces, respectively. $\rho$ and $p$ represent the density and pressure, respectively, while $\nu$ and $\mu=\rho \nu$ correspond to the kinematic and dynamic viscosities of the fluid, respectively. $\phi$ is the passive scalar variable, which is the temperature field $T$ in the present study (i.e. $\phi=T$) and $D_\phi$ is the coefficient of diffusivity. Equations~(1a)-(1c) represent the  axisymmetric NS equations for the axial and radial components of the velocity field in the meridian plane. The structure of the evolution equations for the azimuthal momentum $(\rho u_\theta)$ and  the scalar field ($\phi$) given in Eqs.~(1d) and (1e) respectively, is similar in form, viz., advection-diffusion equation with a source, and hence they can be solved using the same numerical procedures.\par
 In order to represent the above macroscopic equations in cylindrical coordinates in a set of pseudo-2D Cartesian forms, we apply the following coordinate/variable transformations:
 \begin{eqnarray}
(r,z)	\longmapsto (y,x), \quad (u_r,u_z)	\longmapsto (u_y,u_x),\quad \rho u_\theta \longmapsto \psi.
\label{eq:2}
\end{eqnarray}
Then, the resulting equations in pseudo-Cartesian forms involve additional terms when compared to the standard flow and thermal transport equations in 2D, which can be regarded as geometric source terms. The latter will be introduced via a symmetric operator splitting technique in the respective cascaded LB formulation in the following. Thus, the mass and momentum equations for the fluid motion in the meridian plane (Eqs.(1a)-(1c)) can be written in pseudo-2D Cartesian forms as
 \begin{subequations}
\begin{flalign}
  \partial_t \rho+\partial_y(\rho u_y)+\partial_x(\rho u_x)=&M^A&,\label{eq:3a}\\
  \partial_t({\rho u_x})+\partial_r({\rho u_x^2})+\partial_y({\rho u_xu_y})=&-\partial_x p+\partial_x[2\mu \partial_xu_x]+\partial_y[\mu (\partial_yu_x+\partial_xu_y)]&\nonumber\\&+F^A_x+F^b_x,& \label{eq:3b}\\
\partial_t({\rho u_y})+\partial_x({\rho u_xu_y})+\partial_y({\rho u^2_y})=&-\partial_y p+\partial_x[\mu(\partial_xu_y+\partial_yu_x)]+\partial_y[2\mu \partial_yu_y]&\nonumber\\&+F^A_y+F^b_y,&\label{eq:3c}
\end{flalign}
\end{subequations}
where the geometric mass source $M^A$ and the momentum source vector $\bm F^A=(F_x^A,F_y^A)$ can be represented as
\begin{subequations}
\begin{flalign}
  &M^A=-\rho\dfrac{u_y}{y},\label{eq:4a}\\
  &F^A_x=-\rho\dfrac{u_xu_y}{y}+\mu\dfrac{(\partial_xu_y+\partial_yu_x)}{y},\label{eq:4b}\\
&F^A_y=\dfrac{\psi^2}{\rho y}-\rho\dfrac{u_y^2}{y}+2\mu\dfrac{\partial_yu_y}{y}-2\mu \dfrac{u_y}{y^2}.\label{eq:4c}
\end{flalign}
\end{subequations}
Then, the total force $\bm F=(F_x,F_y)$ in this approach becomes
\begin{eqnarray}
 F_x=F_x^A+F_x^b, \quad   F_y=F_y^A+F_y^b.\label{eq:5}
\end{eqnarray}
 Here, the body force $\bm F^b=(F_x^b,F_y^b)$  could be a volumetric force such as the buoyancy force or the Lorentz force. Similarly, the azimuthal momentum equation for $\psi=\rho u_\theta$ can be written as
\begin{equation}
{\partial_t} \psi+{\partial_x}(u_x\psi)+{\partial_y}(u_y\psi)=D_\psi(\partial_x^2 \psi+\partial_y^2 \psi)+S_\psi,\label{eq:6}
\end{equation}
where the corresponding geometric source term $S_\psi$ can be expressed as
\begin{equation}
S_\psi=-\dfrac{2u_y \psi}{y}+\dfrac{\mu}{y}\partial_y (\dfrac{\psi}{\rho})-\dfrac{\nu \psi}{y^2},\label{eq:7}
\end{equation}
and $D_\psi$ is the coefficient of diffusivity, which is equal to the kinematic viscosity of the fluid $\nu$, i.e. $D_\psi=\nu$. Finally, the axisymmetric advection-diffusion equation for the scalar, i.e., temperature field ($\phi=T$) in the pseudo-2D cartesian coordinate system reads as
\begin{equation}
{\partial_t} \phi+{\partial_x}(u_x\phi)+{\partial_y}(u_y\phi)=\partial_x(D_\phi\partial_x \phi)+\partial_y(D_\phi\partial_y \phi)+S_\phi,\label{eq:8}
\end{equation}
where the source term $S_\phi$ is given as
\begin{equation}
S_\phi=-\dfrac{u_y\phi}{y}+\dfrac{D_\phi}{y}\partial_y \phi.\label{eq:9}
\end{equation}
Our goal, then, is to represent the evolution of the axial and radial momentum components along with density (Eqs.(3)-(5)) using a distribution function $f_\alpha$, azimuthal momentum (Eqs.(6)-(7)) using another distribution function $g_\alpha$, and the scalar temperature field (Eqs.(8)-(9)) using a third distribution function $h_\alpha$. We use a D2Q9 lattice for $f_\alpha$, while for both $g_\alpha$ and $h_\alpha$, a D2Q5 lattice would suffice since to represent advection-diffusion type equations, the lattice is required to satisfy only a lower degree of symmetry than the lattice used
for the Navier-Stokes equations. In each case, a cascaded LB scheme based on a symmetric operator splitting will be constructed in the following.
\subsection{Cascaded LB scheme for axial and radial velocity fields: operator splitting for mass and momentum source terms}
In order to consistently include the geometric mass and momentum sources along with any external body force given in Eqs.(4) and (5) in a cascaded LB scheme, we will employ s symmetric operator splitting strategy around its collision term~\cite{hajabdollahi2018Symmetrized}. First, we define the following components of the particle velocities for the D2Q9 lattice:
\begin{subequations}
\begin{eqnarray}
&\ket{e_{ x}} =\left(     0,     1,    0,     -1,     0,  1,-1,-1,1  \right)^\dag,\label{eq:10a}\\
&\ket{e_{ y}} =\left(     0,     0,     1,     0,    -1,1,1,-1,-1
\right)^\dag,\label{eq:10b}
\end{eqnarray}
\end{subequations}
where $\dag$ is the transpose operator and their components for any particle direction $\alpha$ are denoted by $e_{\alpha x}$ and $e_{\alpha y}$, where $\alpha=0,1,\cdots 8$. We also need the following 9-dimensional vector
\begin{eqnarray}
&\ket{1} =\left(     1,     1,    1,     1,     1,  1,1,1,1  \right)^\dag\label{eq:11}
\end{eqnarray}
whose inner product with the distribution function $f_\alpha$ defines its zeroth moment. Here, and in the following, we have used the standard Dirac's bra-ket notation to represent the vectors. The corresponding nine orthogonal basis vectors may be represented by (e.g.~\cite{premnath2009incorporating}):
\begin{eqnarray}
{{K}_{0}}=\ket{1},\quad{{K}_{1}}=\ket{e_{ x}},\quad{{K}_{2}}=\ket{e_{ y}}, {{K}_{3}}=3\ket{e_{ x}^2+e_{ y}^2}-4\ket{1},\nonumber \\
{{K}_{4}}=\ket{e_{ x}^2-e_{ y}^2},\quad
{{K}_{5}}=\ket{e_{ x}e_{ y}},\quad
{{K}_{6}}=-3\ket{e_{ x}^2e_{ y}}+2\ket{e_{ y}},\nonumber \\
{{K}_{7}}=-3\ket{e_{ x}e_{ y}^2}+2\ket{e_{ x}},\quad
{{K}_{8}}=9\ket{e_{ x}^2e_{ y}^2}-6\ket{e_{ x}^2+e_{ y}^2}+4\ket{1}.\label{eq:12}
\end{eqnarray}
Here and henceforth, symbols such as $\ket{e_{ x}^2e_{ y}}=\ket{e_{ x}e_{ x}e_{ y}}$ denote a vector that result from the element wise vector multiplication of vectors $\ket{e_{ x}}$, $\ket{e_{ x}}$ and $\ket{e_{ y}}$. The above set of vectors can be organized by the following orthogonal matrix
\begin{equation}
\tensr{K}=\left[{{K}_{0}},{{K}_{1}},{{K}_{2}},{{K}_{3}},{{K}_{4}},{{K}_{5}},{{K}_{6}},{{K}_{7}},{{K}_{8}}\right],
\label{eq:13}
\end{equation}
which maps changes of moments under collisions due to a cascaded central moment relaxation back to changes in the distribution function (see below). As the cascaded collision operator is built on the moment space, we first define the central moments and raw moments of order ($m+n$) of the distribution function $f_\alpha$ and its equilibrium $f_\alpha^{eq}$ as
\begin{eqnarray}
\left( {\begin{array}{*{20}{l}}
{{{\hat \kappa }_{{x^m}{y^n}}}}\\
{\hat {\kappa} _{{x^m}{y^n}}^{eq}}
\end{array}} \right) = \sum\limits_\alpha  {\left( {\begin{array}{*{20}{l}}
{{f_\alpha }}\\
{f_\alpha ^{eq}}
\end{array}} \right)} {{(e_{\alpha x}-u_x)}^m}{{(e_{\alpha y}-u_y)}^n},
\label{eq:14}
\end{eqnarray}
and
\begin{eqnarray}
\left( {\begin{array}{*{20}{l}}
{{{\hat \kappa }_{{x^m}{y^n}}}}^{'}\\
{\hat {\kappa} _{{x^m}{y^n}}^{eq'}}
\end{array}} \right) = \sum\limits_\alpha  {\left( {\begin{array}{*{20}{l}}
{{f_\alpha }}\\
{f_\alpha ^{eq}}
\end{array}} \right)} {e_{\alpha x}^m}{e_{\alpha y}^n},
\label{eq:15}
\end{eqnarray}
respectively. Here and in what follows, the prime ($'$) symbols denote various raw moments. The central moments of the equilibrium are constructed to be equal to those for the Maxwellian, which then serve as attractors during the cascaded collision represented as a relaxation process~\cite{geier2006cascaded}. In the following, an operator splitting based cascaded LB scheme will be constructed to solve Eqs.(3)-(5). First, we represent the solution of the mass and momentum equations in the meridian plane (Eq.(3)) without the respective source terms (i.e. $M_A, F_x^A,F_x^b,F_y^A,F_y^b$) by means of the evolution of the distribution function $f_\alpha$ using the usual collision and streaming steps ($\tensr C$ and $\tensr S$, respectively) as
\begin{subequations}
\begin{eqnarray}
 &\textbf{Step}\, \tensr C:\quad f^{p}_{\alpha}=f_{\alpha}+(\tensr K\cdot \widehat {\mathbf p})_{\alpha},
\label{eq:16a}\\
&\textbf{Step}\, \tensr S:\quad f_{\alpha}(\bm x, t)=f^{p}_{\alpha}(\bm x-\bm e_{\alpha}\Delta t,t),
\label{eq:16b}
\end{eqnarray}
\end{subequations}
 where $\bm  e_\alpha=(\bm  e_{\alpha x}, \bm  e_{\alpha y})$, $\Delta t$ is the time step, $f_\alpha^p$ is the post-collision distribution function at a location $\bm x$ and time $t$. $\mathbf {\widehat p}=(\widehat p_0,\widehat p_1,\widehat p_2\dots \widehat p_8)$ denotes the changes of different moments under collision based on the relaxation of central moments to their equilibria in a cascaded fashion \cite{geier2006cascaded}. With the mass and momentum being conserved during collision $\widehat p_0=\widehat p_1=\widehat p_2=0$, and the changes in the higher order non-conserved moments are given by (\cite{geier2006cascaded,asinari2008generalized,premnath2009incorporating})
\begin{align}
\widehat{p}_3&=\frac{\omega_3}{12}\left\{ {2}{c_s^2}\rho+{\rho(u_x^2+u_y^2)}
-(\widehat{{\kappa}}_{xx}^{'}+\widehat{{\kappa}}_{yy}^{'})
\right\}, \nonumber \\
\widehat{p}_4&=\frac{\omega_4}{4}\left\{{\rho(u_x^2-u_y^2)}
-(\widehat{{\kappa}}_{xx}^{'}-\widehat{{\kappa}}_{yy}^{'})
\right\}, \nonumber \\
\widehat{p}_5&=\frac{\omega_5}{4}\left\{{\rho u_x u_y}
-\widehat{{\kappa}}_{xy}^{'}
\right\},\nonumber \\
\widehat{p}_6&=\frac{\omega_6}{4}\left\{2\rho u_x^2 u_y+\widehat{{\kappa}}_{xxy}^{'}
              -2u_x\widehat{{\kappa}}_{xy}^{'}-u_y\widehat{{\kappa}}_{xx}^{'}
              \right\}-\frac{1}{2}u_y(3\widehat{p}_3+\widehat{p}_4)-2u_x\widehat{p}_5,\nonumber \\
\widehat{p}_7&=\frac{\omega_7}{4}\left\{2\rho u_x u_y^2+\widehat{{\kappa}}_{xyy}^{'}
              -2u_y\widehat{{\kappa}}_{xy}^{'}-u_x\widehat{{\kappa}}_{yy}^{'}
              \right\}-\frac{1}{2}u_x(3\widehat{p}_3-\widehat{p}_4)-2u_y\widehat{p}_5,\nonumber \\
\widehat{p}_8&=\frac{\omega_8}{4}\left\{{c_s^4}\rho+3\rho u_x^2 u_y^2-\left[\widehat{{\kappa}}_{xxyy}^{'}
                                 -2u_x\widehat{{\kappa}}_{xyy}^{'}-2u_y\widehat{{\kappa}}_{xxy}^{'}
                                 +u_x^2\widehat{{\kappa}}_{yy}^{'}+u_y^2\widehat{{\kappa}}_{xx}^{'}\right.\right.
                                 \nonumber \\
                                 &\left.\left.+4u_xu_y\widehat{{\kappa}}_{xy}^{'}
                                 \right]
                                  \right\}-2\widehat{p}_3-\frac{1}{2}u_y^2(3\widehat{p}_3+\widehat{p}_4)
                                  -\frac{1}{2}u_x^2(3\widehat{p}_3-\widehat{p}_4)\nonumber\\
                                  &-4u_xu_y\widehat{p}_5-2u_y\widehat{p}_6
                                  -2u_x\widehat{p}_7.\label{eq:17}
\end{align}
Here, $\omega_3,\omega_4.\cdots \omega_8$ are relaxation parameters, where $\omega_3,\omega_4$ and $\omega_5$ are related to the bulk and shear viscosities and the other $\omega_i$ influence the numerical stability of the method. In particular, the bulk viscosity is given by $\xi=c_s^2(\frac{1}{\omega_3}-\frac{1}{2})\Delta t$ and the shear viscosity by $\nu=c_s^2(\frac{1}{\omega_j}-\frac{1}{2})/ \Delta t$, where $j=4,5$, and $c_s^2=c^2/3,$  where $c=\Delta x/\Delta t$. In this work, we consider the lattice units, where $\Delta x=\Delta t=1$ and hence the speed of sound $c_s=1/\sqrt{3} $, and the higher order relaxation parameters $\omega_6$, $\omega_7$ and $\omega_8$ are set to unity for simplicity. After the streaming step (see Eq.(16b)), the output density field and the velocity field components (designated with a superscript $"o"$) as the zeroth and first moments of $f_\alpha$ , respectively:
\begin{eqnarray}
&\rho^o=\sum_{\alpha=0}^{8}f_\alpha, \quad \rho^ou_x^o=\sum_{\alpha=0}^{8}f_\alpha e_{\alpha x}, \quad \rho^ou_y^o=\sum_{\alpha=0}^{8}f_\alpha e_{\alpha y}
\label{eq:18}
\end{eqnarray}

 We then introduce the influence of the mass source $M_A$ in Eq.~(3a) and the momentum sources $F_x^A=F_x^A+F_x^b$ and $F_y=F_y^A+F_y^b$ in Eqs.~(3b) and (3c), respectively, as the solution of the following two sub problems, referred to as the mass source step $\tensr M$ and momentum source step $\tensr F$, respectively:
\begin{subequations}
\begin{eqnarray}
&\text{Step}\,  \tensr M:\,  \partial_t \rho=M^A,\label{eq:19a} \\
&\text{Step}\,  \tensr F:\, \partial_t(\rho \bm u)=\bm F=\bm F^A+\bm F^b,
\label{eq:19b}
\end{eqnarray}
\end{subequations}
where $\bm u=(u_x,u_y)$ and $\bm F_A=(F_x^A,F_y^b)$ etc. In our previous work \cite{hajabdollahi2018Symmetrized}, we constructed a symmetric operator splitting based approach to incorporate a single momentum source in a cascaded LB method. In the present work, we further extend this approach to symmetric splitting of multiple operators related to mass and momentum sources. In other words, we perform two symmetric steps of half time steps of length $\Delta t/2$ of $\tensr M$ and $\tensr F$, one before and the other after the collision step. The overall symmetrized operator splitting based cascaded LB algorithm implementing all the four operators ($\tensr C,\tensr S,\tensr M$ and $\tensr F$) during the time interval $[t,t+\Delta t]$ may be written as
 \begin{eqnarray}
 {f_{\alpha}}(\bm x,t+\Delta t)= \tensr M^{1/2}\, \tensr F^{1/2}\, \tensr C\, \tensr F^{1/2}\,  \tensr M^{1/2}\, \tensr S\,{f_{\alpha}}(\bm x,t),
\label{eq:20}
\end{eqnarray}
where $\tensr M^{1/2}$ and $\tensr F^{1/2}$ represent solving Eqs.~(19a) and (19b), respectively, over time step $\Delta t/2$. Both of these steps introduce the effect of geometric mass and momentum source and the body forces directly in the momentum space.\par
Solving Eqs.~(19a) and (19b) for the first part of symmetric sequence needed in Eq.(20) yields $\rho-\rho^o=M^A\frac{\Delta t}{2}$, $\rho u_x-\rho u_x^o=F_x\frac{\Delta t}{2}$ and $\rho u_y-\rho u_y^o=F_y\frac{\Delta t}{2}$. Thus, we have
\begin{subequations}
\begin{flalign}
&\text{Pre-collision Mass Source Step}\,\tensr M^{1/2}:\,\,  &\rho=\rho^o+ M^A\frac{\Delta t}{2}\label{eq:21a}\\
&\text{Pre-collision Momentum Source  Step}\,\tensr F^{1/2}:\,\,  &\rho u_x=\rho u_x^o+F_x\frac{\Delta t}{2},
\label{eq:21b}\\
&\; \; \;&\rho u_y=\rho u_y^o+F_y\frac{\Delta t}{2},
\label{eq:21c}
\end{flalign}
\end{subequations}
where $ M^A, F_x$ and $F_y$ are given in Eqs.~(4a)-(4c) and (5). Based on Eq.~(20), the next step is the collision step, which is performed using the updated density and velocity fields ($\rho,u_x,u_y$) given in Eqs.~(21a)-(21c) and then determining the change of moments under collision ${\widehat p_\beta}~(\beta=3,4\dots 8)$  using Eq.~(17). Then, implementing the other part of the symmetrized mass and momentum steps with using a half time step to solve Eqs.~(19a) and (19b), we obtain the target density and velocity field after collision represented as $(\rho^p,u_x^p,u_y^p)$ via $\rho^p-\rho= M^A\frac{\Delta t}{2}, \; \rho u_x^p-\rho u_x=F_x\frac{\Delta t}{2}$ and $\rho u_y^p-\rho u_y=F_y\frac{\Delta t}{2}$. Thus, we have
\begin{subequations}
\begin{flalign}
\text{Post-collision Momentum Source  Step}\,\tensr F^{1/2}:  &\rho u_x^p=\rho u_x+F_x\frac{\Delta t}{2}\nonumber \\
\; \; & u_y^p=\rho u_y+F_y\frac{\Delta t}{2},\label{eq:22b}\\
\text{Post-collision Mass Source  Step}\,\tensr M^{1/2}:\,\,  &\rho^p=\rho + M^A\frac{\Delta t}{2},
\label{eq:22c}
\end{flalign}
\end{subequations}
By rewriting the above results for the post-collision source steps in terms of the output density $\rho^o$ and velocity field $\bm u^o=(u_x^o,u_y^o)$ via Eqs.~(21a)-(21c), we get
\begin{equation}
\rho^p=\rho^o+{M^A}\Delta t,\quad \rho u^p_x=\rho u^o_x+{F_x}\Delta t,\quad \rho u^p_y=\rho u^o_y+{F_y}\Delta t.
\label{eq:23}
\end{equation}
To effectively design the post-collision distribution function $f_\alpha^p$ in the cascaded LB scheme so that Eq.(23) is precisely satisfied, we consider $ f^{p}_{\alpha}=f_{\alpha}+(\tensr K\cdot \widehat {\mathbf p})_{\alpha}$ and taking its zeroth and first moments, we obtain
\begin{subequations}
\begin{eqnarray}
 &\rho^p=\Sigma_{\alpha}f^p_{\alpha}=\Sigma_{\alpha}f_{\alpha}+\Sigma_{\beta}{\braket{{K_{\beta}}|{1}}} \widehat p_{\beta},
\label{eq:24a}\\
 &\rho u^p_x=\Sigma_{\alpha}f^p_{\alpha}e_{\alpha x}=\Sigma_{\alpha}f_{\alpha}e_{\alpha x}+\Sigma_{\beta}{\braket{{K_{\beta}}|{e_{ x}}}} \widehat p_{\beta},
\label{eq:24b}\\
 &\rho u^p_y=\Sigma_{\alpha}f^p_{\alpha}e_{\alpha y}=\Sigma_{\alpha}f_{\alpha}e_{\alpha y}+\Sigma_{\beta}{\braket{{K_{\beta}}|{e_{ y}}}} \widehat p_{\beta}.
\label{eq:24c}
\end{eqnarray}
\end{subequations}
Since the orthogonal basis vectors $ \ket{K_{\beta}}$ given in Eq.~(12) satisfy $
 \Sigma_{\beta}{\braket{{K_{\beta}}|1}}=9 \widehat {p}_0, \Sigma_{\beta}{\braket{{K_{\beta}}|{e_{x}}}}=6 \widehat {p}_1,\Sigma_{\beta}{\braket{{K_{\beta}}|{e_{ y}}}}=6 \widehat {p}_2,$  Eqs.~(24a)-(24c) become
 \begin{equation}
\rho^p=\rho^o+9 \widehat {p}_0, \quad
\rho u^p_x=\rho u^o_x+6 \widehat {p}_1, \quad \rho u^p_y=\rho u^o_y+6 \widehat {p}_2.
\label{eq:25}
\end{equation}
Comparing Eqs.(23) and (25), it follows that the change of the zeroth moment $(\widehat {p}_0)$ and the first moments $(\widehat {p}_1$ and $\widehat {p}_2)$ due to mass and momentum source can be written as
\begin{equation}
\widehat {p}_0=\frac{M^A}{9}\Delta t, \quad\widehat {p}_1=\frac{F_x}{6}\Delta t, \quad \widehat {p}_2=\frac{F_y}{6}\Delta t.
\label{eq:26}
\end{equation}
where $M_A$ follows from Eq.~(4a), $F_x$ and $F_y$ are given in Eq.~(5) and (4b)-(4c). These expressions effectively provide the desired post-collision states of the distribution function, i.e. $f_\alpha^p$ due to mass and momentum sources. Thus, finally expanding $(\tensr K\cdot \widehat {\mathbf p})_{\alpha}$ in Eq.~(16a), the components of the post-collision distribution functions read as
\begin{eqnarray}
f^p_0&=&{f}_{0}+\left[\widehat{p}_0-4(\widehat{p}_3-\widehat{p}_8)\right], \nonumber \\
f^p_1&=&{f}_{1}+\left[\widehat{p}_0+\widehat{p}_1-\widehat{p}_3+\widehat{p}_4    +2(\widehat{p}_7-\widehat{p}_8)\right], \nonumber \\
f^p_2&=&{f}_{2}+\left[\widehat{p}_0+\widehat{p}_2-\widehat{p}_3-\widehat{p}_4
+2(\widehat{p}_6-\widehat{p}_8)\right], \nonumber \\
f^p_3&=&{f}_{3}+\left[\widehat{p}_0-\widehat{p}_1-\widehat{p}_3+\widehat{p}_4
-2(\widehat{p}_7+\widehat{p}_8)\right],\nonumber \\
f^p_4&=&{f}_{4}+\left[\widehat{p}_0-\widehat{p}_2-\widehat{p}_3-\widehat{p}_4
-2(\widehat{p}_6+\widehat{p}_8)\right], \nonumber \\
f^p_5&=&{f}_{5}+\left[\widehat{p}_0+\widehat{p}_1+\widehat{p}_2+2\widehat{p}_3
+\widehat{p}_5-\widehat{p}_6-\widehat{p}_7+\widehat{p}_8\right], \nonumber \\
f^p_6&=&{f}_{6}+\left[\widehat{p}_0-\widehat{p}_1+\widehat{p}_2+2\widehat{p}_3
-\widehat{p}_5-\widehat{p}_6+\widehat{p}_7+\widehat{p}_8\right], \nonumber \\
f^p_7&=&{f}_{7}+\left[\widehat{p}_0-\widehat{p}_1-\widehat{p}_2+2\widehat{p}_3
+\widehat{p}_5+\widehat{p}_6+\widehat{p}_7+\widehat{p}_8\right], \nonumber \\
f^p_8&=&{f}_{8}+\left[\widehat{p}_0+\widehat{p}_1-\widehat{p}_2+2\widehat{p}_3
-\widehat{p}_5+\widehat{p}_6-\widehat{p}_7+\widehat{p}_8\right]. \label{eq:27}
\end{eqnarray}
where $\widehat{p}_0$, $\widehat{p}_1$ and $\widehat{p}_2$ are obtained from Eq.~(26) and $\widehat{p}_3,\widehat{p}_4,\cdots,\widehat{p}_8$ from Eq.~(17)
\subsection{Cascaded LB scheme for azimuthal velocity field:~operator splitting for source term}
We now construct a novel cascaded LB scheme for the solution of the equation of the azimuthal momentum component ($\psi=\rho u_\theta$) given in Eqs.~(6) and (7) using a D2Q5 lattice \cite{hajabdollahi2018Symmetrized}. First, defining the vectors corresponding to particle velocity components and a 5-dimensional vector $\ket{1}$ as
\begin{subequations}
\begin{eqnarray}
&\ket{e_{ x}} =\left(     0,     1,    0,     -1,     0  \right)^\dag,\label{eq:28a}\\
&\ket{e_{y}} =\left(     0,     0,     1,     0,    -1\right)^\dag,\label{eq: 28b}\\
&\ket{1} =\left(     1,     1,     1,     1,    1
\right)^\dag,\label{eq:28c}
\end{eqnarray}
\end{subequations}
where taking the inner product of the distribution function $g_\alpha$ with  $\ket{1}$  defines its zeroth moment. Using these, the five orthogonal basis vectors can be written as
\begin{eqnarray}
{{L}_{0}}=\ket{1},\,
{{L}_{1}}=\ket{e_{ x}},\,
{{L}_{2}}=\ket{e_{ y}},\nonumber \\
{{L}_{3}}=5\ket{e_{ x}^2+e_{ y}^2}-4\ket{1},\,
{{L}_{4}}=\ket{e_{ x}^2-e_{ y}^2},\label{eq:29}
\end{eqnarray}
 which can be grouped together as the following transformation matrix that converts the changes in moments to those in the distribution functions:
 \begin{equation}
\tensr{L}=\left[{{L}_{0}},{{L}_{1}},{{L}_{2}},{{L}_{3}},{{L}_{4}}\right].
\label{eq:30}
\end{equation}

In order to design a cascaded collision operator to solve for the azimuthal momentum, which acts as a passive scalar field $\psi=\rho u_\theta$ described by on advection-diffusion equation under the action of a local source term (Eqs.~(6) and (7)), we define the following central moments and raw moments of the distribution function $g_\alpha$ and its equilibrium $g_\alpha^{eq}$ as
\begin{eqnarray}
\left( \begin{array}{l}
{{\hat \kappa }_{{x^m}{y^n}}}^{\psi}\\
\hat \kappa _{{x^m}{y^n}}^{eq,\psi}
\end{array} \right) = \sum\limits_\alpha  {\left( \begin{array}{l}
{g_\alpha }\\
g_\alpha ^{eq}\\
\end{array} \right)} {({e_{\alpha x}} - {u_x})^m}{({e_{\alpha y}} - {u_y})^n},\label{eq:31}
\end{eqnarray}
and
\begin{eqnarray}
\left( {\begin{array}{*{20}{l}}
{{{\hat \kappa }_{{x^m}{y^n}}}}^{\psi'}\\
{\hat {\kappa} _{{x^m}{y^n}}^{{eq,\psi'}}}
\end{array}} \right) = \sum\limits_\alpha  {\left( {\begin{array}{*{20}{l}}
{{g_\alpha }}\\
{g_\alpha ^{eq}}
\end{array}} \right)} {e_{\alpha x}^m}{e_{\alpha y}^n}.
\label{eq:32}
\end{eqnarray}
 respectively. The central moments of the equilibrium $\hat \kappa _{{x^m}{y^n}}^{eq,\psi}$ are devised  be equal to those for the Maxwellian after replacing the density with the scalar field in its expression. Then the cascaded collision step is written in terms of relaxation of different central moments to their equilibria. Similar to the previous section, a symmetrized operator split scheme will now be developed to solve Eqs.~(6) and (7) in the cascaded LB formulation. First, we represent the solution of Eq.~(6) without the source term (Eq.~(7)) through the collision and streaming steps of the distribution function $g_\alpha$ as
\begin{subequations}
\begin{eqnarray}
 &&\textbf{Step}\, \tensr C:\quad g^{p}_{\alpha}=g_{\alpha}+(\tensr L\cdot \widehat {\mathbf q})_{\alpha},
\label{eq:33a}\\
&&\textbf{Step}\, \tensr S:\quad g_{\alpha}(\bm x, t)=g^{p}_{\alpha}(\bm x-\bm e_{\alpha}\Delta t,t).
\label{eq:33b}
\end{eqnarray}
\end{subequations}
 where $g_\alpha^p$ is the post-collision distribution function and $\widehat {\mathbf q}=(\widehat { q}_o,\widehat {q}_1,\cdots \widehat {q}_4)$ represents the changes of different moments under a cascaded collision prescribed as a relaxation process in terms of central moments, which reads as \cite{hajabdollahi2018Symmetrized}
\begin{eqnarray}
\widehat{q}_1&=&\frac{\omega_1^\psi}{2}\left[\psi u_x-{\widehat\kappa}_x^{\psi'}\right],  \nonumber \\
\widehat{q}_2&=&\frac{\omega_2^\psi}{2}\left[\psi u_y-{\widehat\kappa}_y^{\psi'}\right], \nonumber \\
\widehat{q}_3&=&\frac{\omega_3^\psi}{4}\left[2c_{s\psi}^2-({{\widehat\kappa}_{xx}}^{\psi'}+{{\widehat\kappa}_{yy}}^{\psi'})+2(u_x{\widehat\kappa}_x^{\psi'}+u_y{\widehat\kappa}_y^{\psi'})+(u_x^2+u_y^2)\psi\right]+u_x\widehat q_1+u_y\widehat q_2, \nonumber \\
\widehat{q}_4&=&\frac{\omega_4^\psi}{4}\left[-({{\widehat\kappa}_{xx}}^{\psi'}-{{\widehat\kappa}_{yy}}^{'\psi})+2(u_x{\widehat\kappa}_x^{\psi'}-u_y{\widehat\kappa}_y^{\psi'})+(u_x^2-u_y^2)\psi\right]+u_x\widehat q_1-u_y \widehat q_2,
\label{eq:34}
\end{eqnarray}
where $\omega_1^\psi$, $\omega_2^\psi$, $\omega_3^\psi$ and $\omega_4^\psi$ are the relaxation parameters. Since $\psi$ is conserved during collision, $\widehat q_o=0$. The relaxation parameters for the first order moments ($\omega_1^\psi$ and $\omega_2^\psi$) are related to diffusivity $D_\psi=\nu=c_{s \psi}^2 (\frac{1}{\omega_j^\psi}-\frac{1}{2})\Delta t$, j=1,2 where $c_{s \psi}^2$ is a free parameter, which is set to $1/3$. The relaxation parameters for the higher order moments, which influence numerical stability, are taken to be unity in this study. After the streaming step in Eq.(33b), the output passive azimuthal momentum field $\psi^o$ is computed as the zeroth moment of $g_\alpha$ as
\begin{eqnarray}
 \quad \psi^o =\sum_{\alpha}^4 {g}_{\alpha}.
\label{eq:35}
\end{eqnarray}

The source term $S_\psi$, which was eliminated in the above, will now be introduced by appropriately combining its effect after solution of the following such problem:
\begin{eqnarray}
 &\textbf{Step}\, \tensr R:\quad \partial_t \psi=S_\psi
\label{eq:36}
\end{eqnarray}
Its solution will now be combined with the split solution obtained in the absence of the source term in Eqs.(33a) and (33b) via a symmetric operator splitting technique over a time interval $[t,t+\Delta t]$, analogous to that considered in the previous subsection. This can be represented as
\begin{eqnarray}
 {g_{\alpha}}(\bm x,t+\Delta t)=\tensr S\, \tensr R^{1/2}\,\tensr C\, \tensr R^{1/2}{g_{\alpha}}(\bm x,t),
\label{eq:37}
\end{eqnarray}
The pre-collision source step $ \tensr R^{1/2}$ is executed via a solution of Eq.(36) over a duration $\Delta t/2$, which yields $\psi-\psi^o=S_\phi \frac{\Delta t}{2}$, and hence
\begin{eqnarray}
&\text{Pre-collision Source Step}\,\tensr R^{1/2}:\,\,  \psi=\psi^o+S_\psi\frac{\Delta t}{2}\label{eq:38}
\end{eqnarray}
 Based on this updated scalar field, the changes of different moments under collision $\widehat q_\beta,  \beta=1,2,3,4$, given in Eq.(34) can be computed. Similarly, the other part of the source step $\tensr R^{1/2}$ with half time step following collision can be performed by solving Eq.(36), which can be expressed as
\begin{eqnarray}
&\text{Post-collision Source Step}\,  \tensr R^{1/2}: \psi^p=\psi +{S_\psi}\frac {\Delta t} {2}\label{eq:39}
\end{eqnarray}
where $\psi^p$ is the target scalar field after collision. By rewriting it in terms of the output scalar field $\psi^o$  using Eq.~(38), we have
\begin{equation}
\psi^p=\psi^o+{S_\psi}\Delta t.
\label{eq:40}
\end{equation}
In order for the post-collision distribution function $ g^{p}_{\alpha}=g_{\alpha}+(\tensr L\cdot \widehat {\mathbf q})_{\alpha}$ to satisfy Eq.~(40), we write its zeroth moment as
\begin{eqnarray}
 &\psi^p=\Sigma_{\alpha}g^p_{\alpha}=\Sigma_{\alpha}g_{\alpha}+\Sigma_{\beta}{\braket{{L_{\beta}}|1}} \widehat q_{\beta}.
\label{eq:400}
 \end{eqnarray}
Since $\Sigma_{\beta} \braket{{L_{\beta}}|1}q_{\beta}=5\widehat q_o$ via orthogonal of basis vectors (see Eq.(29)), it follows from Eqs.(35) and (41) that $\psi^p=\psi^o+5 \widehat {q}_0$. Comparing this with Eq.(40), we get the change of the zeroth moment $\widehat q_o$ due to the presence of the source term $S_\psi$ as
\begin{equation}
\widehat {q}_0=\frac{S_\psi}{5}\Delta t.
\label{eq:41}
\end{equation}
Finally, the components of the post-collision distribution function in Eq.(33a) can be expressed after expanding $(\tensr L\cdot \widehat {\mathbf q})_{\alpha} $ as
\begin{eqnarray}
{g}_{0}^p&=&{g}_{0}+\left[\widehat{q}_0-4\widehat{q}_3\right],\nonumber \\
 {g}_{1}^p&=&{g}_{1}+\left[\widehat{q}_0+\widehat{q}_1+\widehat{q}_3+\widehat{q}_4\right],\nonumber \\
{g}_{2}^p&=&{g}_{2}+\left[\widehat{q}_0+\widehat{q}_2+\widehat{q}_3-\widehat{q}_4\right], \nonumber  \\
{g}_{3}^p&=&{g}_{3}+\left[\widehat{q}_0-\widehat{q}_1+\widehat{q}_3+\widehat{q}_4\right],\nonumber \\
 {g}_{4}^p&=&{g}_{4}+\left[\widehat{q}_0-\widehat{q}_2+\widehat{q}_3-\widehat{q}_4\right],
 \label{eq:42}
\end{eqnarray}
where $\widehat{q}_o$ (i.e., the change of the zeroth moment due to source) is given in Eq.~(42) and $\widehat{q}_\beta , \beta=1,2,3,4$ (i.e., the changes of the higher, non-conserved, moments under collision) is obtained from Eq.~(34).
\subsection{Cascaded LB scheme for temperature field:~operator splitting for source term}
As in the previous section, we consider a D2Q5 lattice, and use the orthogonal basis vectors $L_\beta$ and the transformation matrix $\tensr L$ given in Eqs.~(29) and (30), respectively, to design a cascaded LB scheme for the solution of the temperature field $\phi=T$. Its evolution is presented by the advection-diffusion equation with a source term given in Eqs.~(8) and (9). The various central moments and raw moments of the corresponding distribution function $h_\alpha$ and its equilibrium $h_\alpha^{eq}$ are defined as
\begin{eqnarray}
\left( \begin{array}{l}
{{\hat \kappa }_{{x^m}{y^n}}}^{\phi}\\
\hat \kappa _{{x^m}{y^n}}^{eq,\phi}
\end{array} \right) = \sum\limits_\alpha  {\left( \begin{array}{l}
{h_\alpha }\\
h_\alpha ^{eq}\\
\end{array} \right)} {({e_{\alpha x}} - {u_x})^m}{({e_{\alpha y}} - {u_y})^n},\label{eq:43}
\end{eqnarray}
and
\begin{eqnarray}
\left( {\begin{array}{*{20}{l}}
{{{\hat \kappa }_{{x^m}{y^n}}}}^{\phi'}\\
{\hat {\kappa} _{{x^m}{y^n}}^{{eq,\phi'}}}
\end{array}} \right) = \sum\limits_\alpha  {\left( {\begin{array}{*{20}{l}}
{{h_\alpha }}\\
{h_\alpha ^{eq}}
\end{array}} \right)} {e_{\alpha x}^m}{e_{\alpha y}^n}.
\label{eq:44}
\end{eqnarray}
As before, we use the symmetrized operator splitting to include the source term $S_\phi$ in the cascaded LB scheme, which can be presented as :
\begin{eqnarray}
 {h_{\alpha}}(\bm x,t+\Delta t)=\tensr S\, \tensr R^{1/2}\,\tensr C\, \tensr R^{1/2}{h_{\alpha}}(\bm x,t),
\label{eq:45}
\end{eqnarray}
where $\tensr C$ and $\tensr S$ denote the collision and streaming steps, respectively, of $g_\alpha$ used to solve Eq.~(8) (without $S_\phi$)
\begin{subequations}
\begin{eqnarray}
\textbf{Step \tensr C:}\,\, h^{p}_{\alpha}&=&h_{\alpha}+(\tensr L\cdot \widehat {\mathbf r})_{\alpha},
\label{eq:46a}\\
 \textbf{Step \tensr S:}\,\, h_{\alpha}(\bm x, t)&=&h^{p}_{\alpha}(\bm x-\bm e_{\alpha}\Delta t,t).
\label{eq:46b}
\end{eqnarray}
\end{subequations}
Here, $h^{p}_{\alpha}$ is the post-collision distribution function and $\widehat {\mathbf r}=(\widehat r_o,\widehat r_1,\widehat r_2,\widehat r_3,\widehat r_4)$ is the change of different moments under collision, with $\widehat r_o=0$ due to $\phi$ being a collision invariant. In the above, after the streaming step, the solution of the output scalar field $\phi^o$ is computed via the zeroth moment of $h_\alpha$ as
\begin{equation}
\phi^o=\sum_{\alpha=0}^4h_{\alpha}.
\label{eq:47}
\end{equation}
 The operator $\tensr R^{1/2}$ applied twice in Eq.(46) represents the split solution of the scalar field due to the source term of the evolution equation $\partial_t \phi=S_\phi$ before and after collision over a half time step $\Delta t/2$. Thus, the pre-collision source step can be expressed as
\begin{eqnarray}
\text{Pre-collision Source Step}\,\tensr R^{1/2}: \phi= \phi^o+ {\frac {S_\phi}{2}}\Delta t.
\label{eq:48}
\end{eqnarray}
This updated scalar field $\phi$ is then used to compute the changes of different moments under collision $\widehat r_\beta, \beta=1,2,3,4$, which can be written as
\begin{eqnarray}
\widehat{r}_1&=&\frac{\omega_1^{\phi}}{2}\left[\phi u_x-{\widehat\kappa}_x^{\phi'}\right],  \nonumber \\
\widehat{r}_2&=&\frac{\omega_2^{\phi}}{2}\left[\phi u_y-{\widehat\kappa}_y^{\phi'}\right], \nonumber \\
\widehat{r}_3&=&\frac{\omega_3^{\phi}}{4}\left[2c_s^2\phi-({{\widehat\kappa}_{xx}}^{\phi'}+{{\widehat\kappa}_{yy}}^{\phi'})+2(u_x{\widehat\kappa}_x^{\phi'}+u_y{\widehat\kappa}_y^{\phi'})+(u_x^2+u_y^2)\phi\right]+u_x\widehat r_1+u_y\widehat r_2, \nonumber \\
\widehat{r}_4&=&\frac{\omega_4^{\phi}}{4}\left[-({{\widehat\kappa}_{xx}}^{\phi'}-{{\widehat\kappa}_{yy}}^{'\phi})+2(u_x{\widehat\kappa}_x^{\phi'}-u_y{\widehat\kappa}_y^{\phi'})+(u_x^2-u_y^2)\phi\right]+u_x\widehat r_1-u_y\widehat r_2,
\label{eq:49}
\end{eqnarray}
where the relaxation parameters $\omega_1^{\phi}$ and $\omega_2^{\phi}$ are related to the thermal diffusivity $D_\phi$ via $D_\phi=c_{s\phi}^2(\frac{1}{\omega_j^\phi}-\frac{1}{2})\Delta t, j=1,2$, where $c_{s\phi}^2=\frac{1}{3}$ and $\omega_3^{\phi}=\omega_4^{\phi}=1$ in this work. Following this, the post-collision source step $\tensr R^{1/2}$ can be represented as
\begin{equation}
\text{Post-collision Source Step}\,\tensr R^{1/2}: \phi^p= \phi+ {\frac {S_\phi}{2}}\Delta t
\label{eq:50}
\end{equation}
where $\phi^p$ is the target scalar field following collision, which via Eq.(49) reads as $\phi^p=\phi^o+S_\phi \Delta t$. The post-collision distribution function $h^{p}_{\alpha}=h_{\alpha}+(\tensr L\cdot \widehat {\mathbf r})_{\alpha}$ can be made to satisfy this condition using Eq. (48) and using $\Sigma_{\beta} \braket{{L_{\beta}}|1}\widehat{r}_\beta=5\widehat{r}_0$ after taking its zeroth moment, i.e. $\phi^p=\sum_{\alpha}{h^p_{\alpha}}$. This provides the following zeroth moment change due to $S_\phi$ after collision
\begin{equation}
\widehat{r}_0=\frac{S_{\phi}}{5}\Delta t.
\label{eq:51}
\end{equation}
Finally, the post-collision distribution function ${h}_{\alpha}^p$ can be explicitly written after expanding $(\tensr L\cdot \widehat {\mathbf r})_{\alpha}$ in Eq.(47a) as follows:
\begin{eqnarray}
{h}_{0}^p&=&{h}_{0}+\left[\widehat{r}_0-4\widehat{r}_3\right],\nonumber \\
 {h}_{1}^p&=&{h}_{1}+\left[\widehat{r}_0+\widehat{r}_1+\widehat{r}_3+\widehat{r}_4\right],\nonumber \\
{h}_{2}^p&=&{h}_{2}+\left[\widehat{r}_0+\widehat{r}_2+\widehat{r}_3-\widehat{r}_4\right], \nonumber  \\
{h}_{3}^p&=&{h}_{3}+\left[\widehat{r}_0-\widehat{r}_1+\widehat{r}_3+\widehat{r}_4\right],\nonumber \\
 {h}_{4}^p&=&{h}_{4}+\left[\widehat{r}_0-\widehat{r}_2+\widehat{r}_3-\widehat{r}_4\right],
 \label{eq:52}
\end{eqnarray}
where $\widehat r_o$ is obtained form Eq.~(52) and $\widehat r_\beta$, $\beta=1,2,3$ and 4, follows from Eq.~(50) due to various non-conserved moment changes under collision.
\section{Results and Discussion}
In this section, the cascaded LB schemes described above will be applied to and studied for different complex flow benchmark problems to validate them for simulations of axisymmetric flows with heat transfer and including rotational/swirling effects. These include the following: (a) Taylor-Couette flow between two rotating
circular cylinders, (b) natural convection in an annulus between two stationary coaxial vertical cylinders, (c) Rayleigh-Benard convection inside vertical cylinder heated at the bottom and cooled at the top, (d) cylindrical cavity flow driven by the motion of the top lid, (e) mixed convection in a slender vertical annulus subjected to the inner cylinder rotation, and  (f) melt flow in a cylinder during Czochralski crystal growth process.
\subsection{Taylor-Couette flow}
As the first test problem, the classical shear-driven circular Couette flow between two circular cylinders is considered \cite{taylor1923stability}. This problem is used to assess the cascaded LB scheme for the azimuthal velocity component $u_\theta$ given in Sec.~2.3, whose evolution is represented by Eqs.~(6) and (7). The radii of the inner and outer cylinders are defined as $R_i$ and $R_o$, respectively. Let the angular velocities of the inner and outer cylinders be $\Omega_i$ and $\Omega_o$, respectively, which induce an azimuthal flow within their annulus gap. The analytical solution for such a cylindrical Couette flow is given in terms of the radial variation of the azimuthal velocity as follows:
\begin{eqnarray*}
u_\theta(r)=A r+\frac{B}{r},
\end{eqnarray*}
where $A=\frac{\Omega_oR_o^2-\Omega_iR_i^2}{R_o^2-R_i^2}$, $B=\frac{(\Omega_i-\Omega_o)R_i^2R_o^2}{R_o^2-R_i^2}$. Here, $r$ is the radial distance from the cylindrical axis. For ease of representation, this can be written in a non-dimensional form as
\begin{eqnarray*}
\frac{u_\theta(r)}{u_o}=\frac{1}{1-\beta^2}[(\kappa-\beta^2)\frac{r}{R_i}+\frac{R_i}{r}(1-\kappa)],
\end{eqnarray*}
where $u_o=\Omega_iR_i$, $\beta$ is the radius ratio given by $\beta=R_i/R_o$ and $\kappa$ denotes the angular velocity ratio, i.e., $\kappa=\Omega_o/\Omega_i.$\par
In our simulation, periodic boundary conditions are applied in the axial direction and the values of the azimuthal velocities at the inner and outer cylinder are prescribed as $u_\theta(r=R_i)=\Omega_iR_i=u_o$ and $u_\theta(r=R_o)=\Omega_oR_o=\dfrac{\kappa}{\beta}u_o$, respectively using the Dirichlet boundary condition implementation scheme associated with the advection-diffusion equation representing the dynamics of $u_\theta$~\cite{yoshida2010multiple}. The outer cylinder radius is resolved by 200 lattice nodes and the lattice location for the inner cylinder fixed using $R_i=\beta R_o$ for different choices of $\beta$. The periodic axial direction is discretized using 3 lattice nodes. The relaxation times in the cascaded LB scheme representing the kinematic shear viscosity are set as $\omega_j=1/\tau, j=4,5$, where $\tau=0.6$, and  $u_o$ is chosen such that the rotational Reynolds number $Re=u_oR_i/\nu$ becomes 5. Figure 1 presents a comparison of the velocity profiles computed using the cascaded LB scheme against the analytical solution at the angular velocity ratio $\kappa=0.1$ for various values of the radius ratio $\beta$ $(\beta=0.103, 0.203, 0.303$ and $0.503)$. It is clear that the agreement between the numerical and analytical solution is very good.
 \begin{figure}[htbp]
\centering
    \subfloat{
        \includegraphics[scale=0.27] {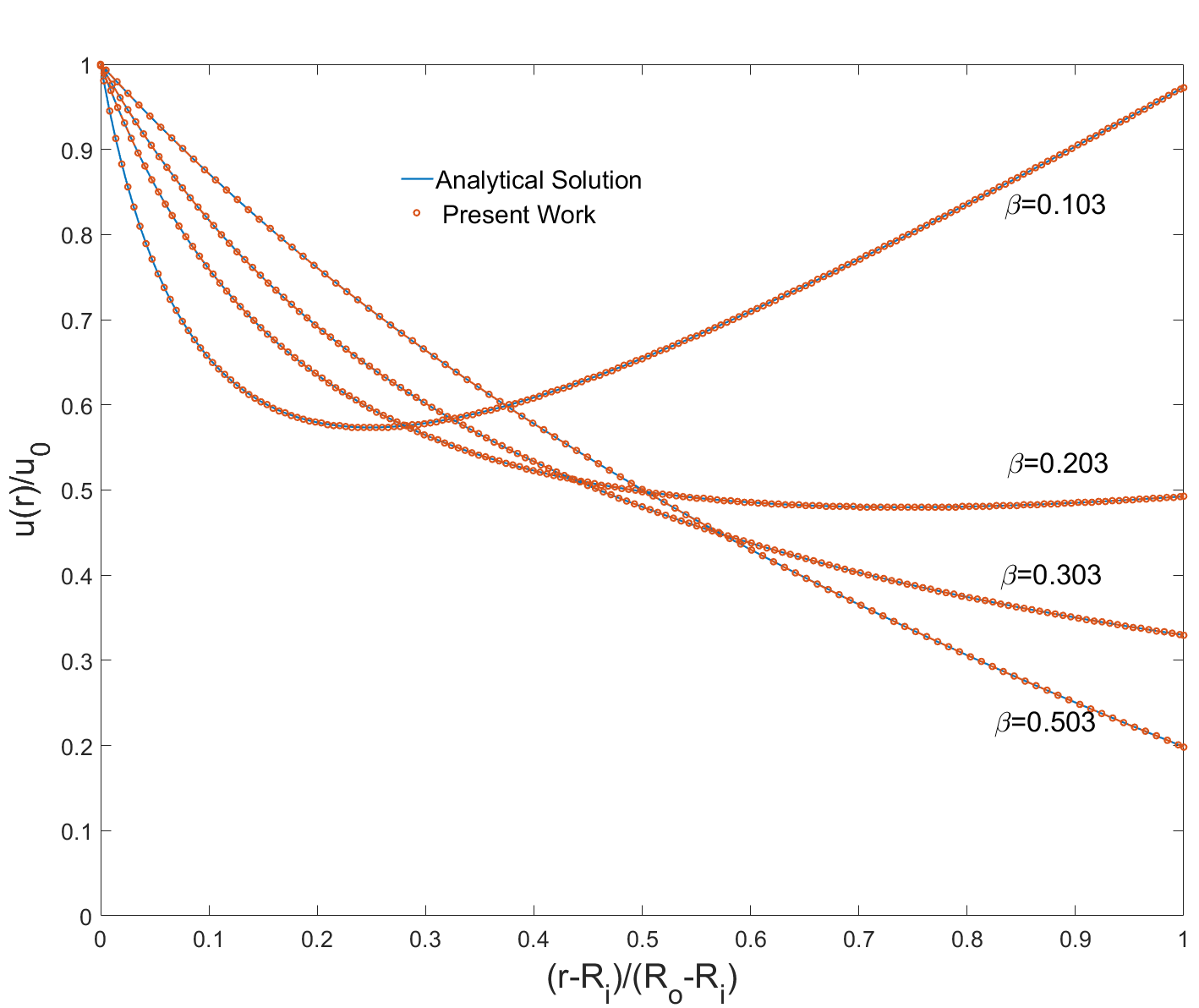}
         }
                    \caption{Comparison between the analytical velocity profile (solid lines) and the cascaded LB solution (symbols) for the Taylor-Couette flow between two circular cylinders at an angular velocity ratio $\kappa=0.1$ and for various values of  the radius ratio $\beta$. }
                     \label{fig:Couette}
    \end{figure}

\subsection{Natural convection in an annulus between two coaxial vertical cylinders}
In order to validate our cascaded LB schemes for axisymmetric flows with heat transfer, we simulate a buoyancy-driven flow between two coaxial stationary cylinders, which is a prototype problem of both fundamental and practical interest. Since the flow field is coupled to the temperature field via the buoyancy force in view of Eqs.~(3a)-(3c), (4a)-(4c), (5), (8) and (9), this problem facilitates a  thorough examination of the efficacy of the coupling between the cascaded LB schemes presented in Sec.~2.2 and 2.4. The schematic of this problem is depicted in Fig~.2, where $R_i$, $R_o$, $H$  and $g$ are the radii of the inner cylinder and outer cylinder, the height of the cylinder and the gravitation acceleration, respectively.
\begin{figure}[htbp]
\centering
    \subfloat{
        \includegraphics[scale=0.4] {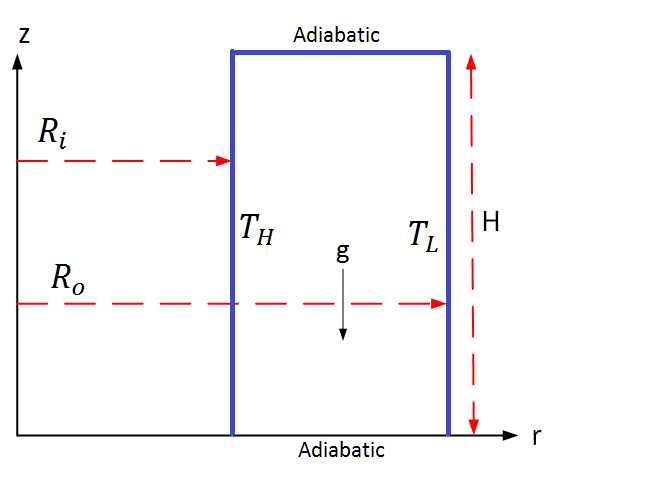}
         }
                    \caption{Schematic illustration of the geometry and boundary conditions for natural convection in a vertical annulus.}
                     \label{fig:Natural}
    \end{figure}

For the velocity field, no-slip boundary conditions are considered on all four walls involving the inner and outer cylindrical surfaces, and top and bottom walls. The inner and outer  walls of the lateral cylindrical side walls are maintained at temperatures of $T_H$ and $T_L$, respectively, where $T_H>T_L$, while the top and bottom walls are considered to be thermally insulated (adiabatic). As a result, this generates a body force due to buoyancy in the axial direction, which under the Boussinesq approximation, can be written as $g\beta (T-T_o)$, where $\beta$ is the thermal expansion coefficient, and $T_o=(T_H+T_L)/2$. This body force component is added to the geometric source terms in Eq.~(5) for $F_x^b$, which then sets up natural convection within the annulus of the axisymmetric  geometry. This thermally driven flow problem is characterized by two dimensionless numbers, viz., the Rayleigh number Ra and Prandtl number $Pr$ defined as
\begin{eqnarray*}
Ra=\frac{g \beta (T_H-T_L)L^3}{\alpha \nu}, \; Pr=\frac{\nu}{\alpha},
\end{eqnarray*}
where $L=R_o-R_i$ is the annual gap serving as the characteristic length, and $\nu$ and $\alpha$ are the kinematic viscosity and thermal diffusivity, respectively. In addition, the geometric parameters influencing this problem are the aspect ratio $H/L$ and the radius ratio $R_o/R_i$, both of which are set to 2 in the present study. The no-slip conditions for the velocity field are implemented using the standard half-way bounce back scheme in the cascaded LB method, while the imposed temperature and no heat flux conditions on the boundaries are represented using the approach presented in \cite{yoshida2010multiple}. All the spatial derivatives needed in the source terms in Eqs.(4b), (4c) and (9) are computed using a central difference scheme, and the computational domain is resolved using a grid resolution of $200\times 200$ in the axial and radial directions, respectively. The characteristic velocity due to natural convection $\sqrt{g\beta(T_H-T_L)R_i}$ is kept small so that the flow can be regarded as incompressible.  We performed simulations at $Pr=0.7$ and $Ra=10^3$, $10^4$ and $10^5$. Figure 3 presents the computed streamlines and isotherms for three different  $Ra=10^3$, $10^4$ and $10^5$.
\begin{figure}[htbp]
\centering
    \subfloat[Re=$10^3$\label{subfig-2:dummy}]{
        \includegraphics[trim={3cm 1cm 7cm 1.3cm},clip,scale=0.3] {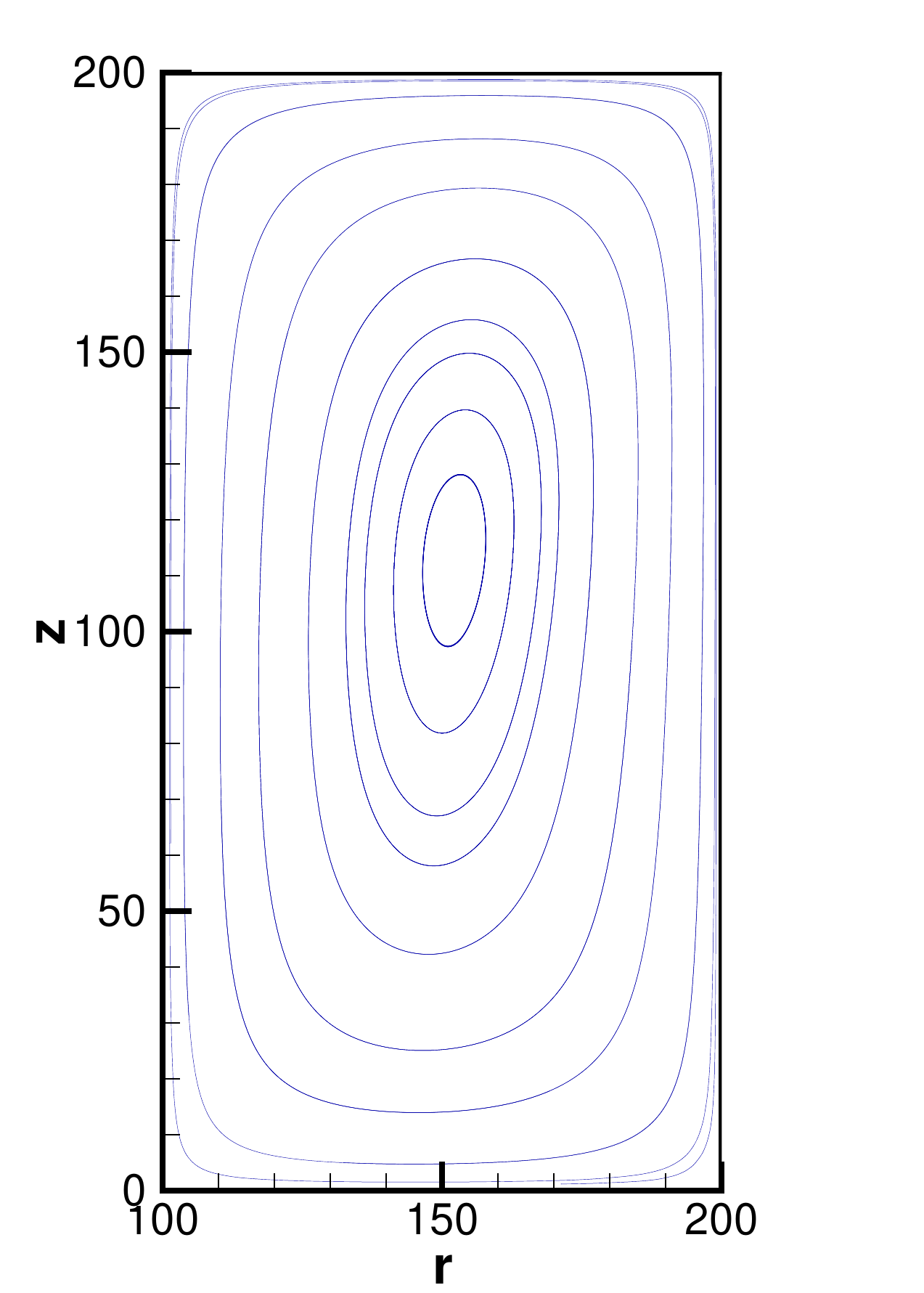}
         }
             \subfloat[Re=$10^4$\label{subfig-2:dummy}]{
       \includegraphics[trim={3cm 1cm 7cm 1.3cm},clip,scale=0.3] {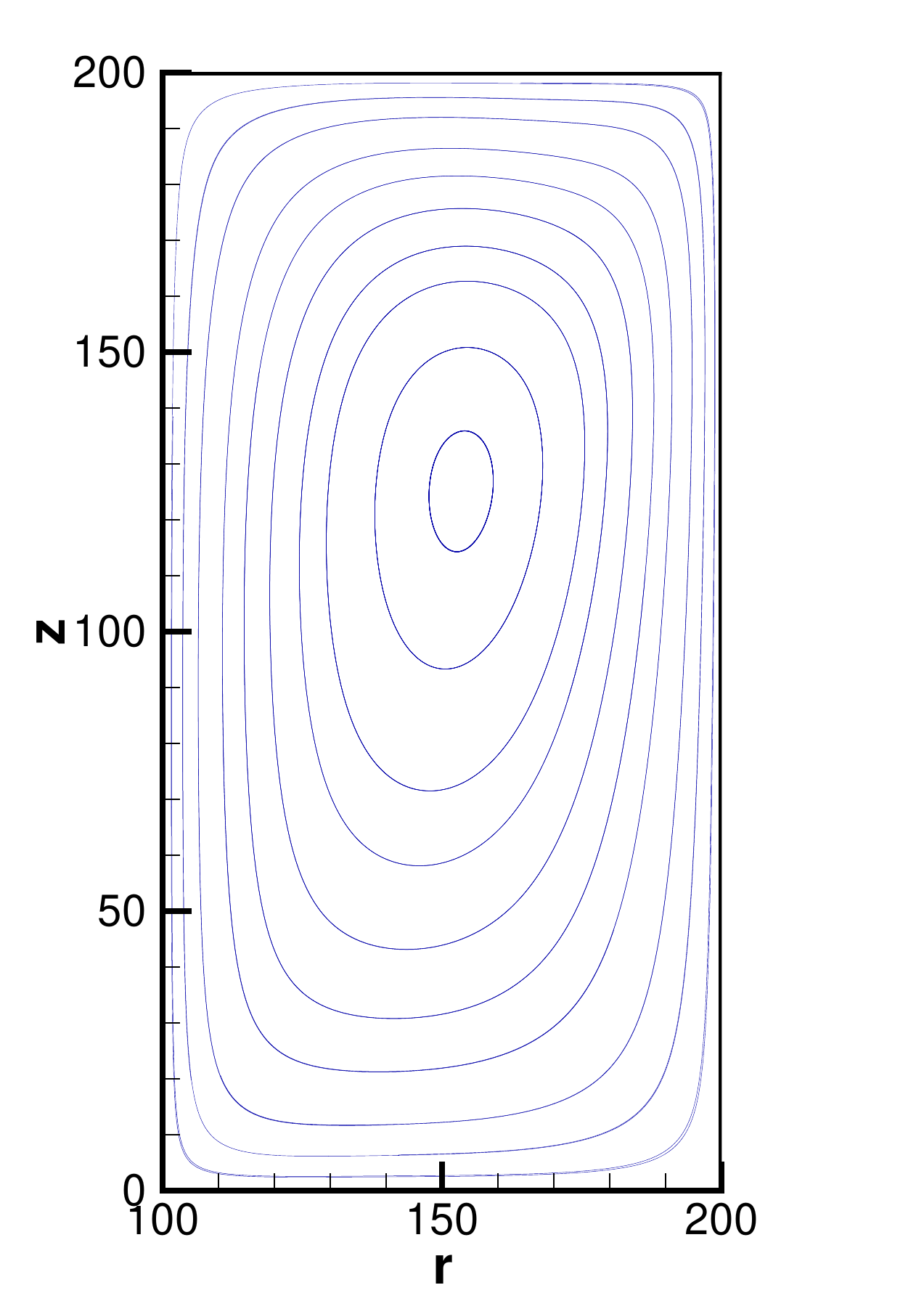}
        }
    \subfloat[Re=$10^5$\label{subfig-2:dummy}]{
        \includegraphics[trim={3cm 1cm 7cm 1.3cm},clip,scale=0.3] {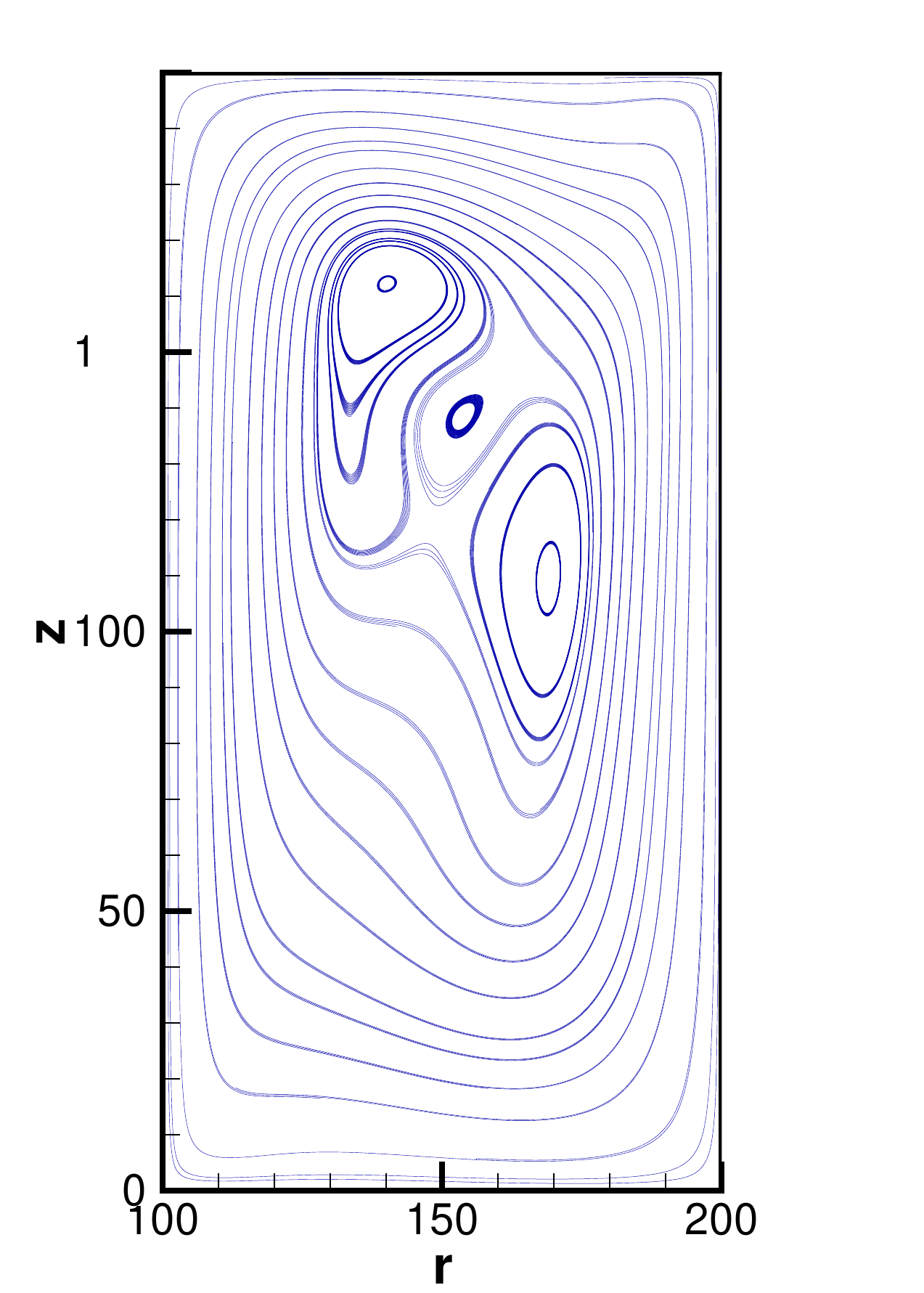}
         }
         \\
          \subfloat[Re=$10^3$\label{subfig-2:dummy}]{
        \includegraphics[trim={3cm 1cm 7cm 1.3cm},clip,scale=0.3] {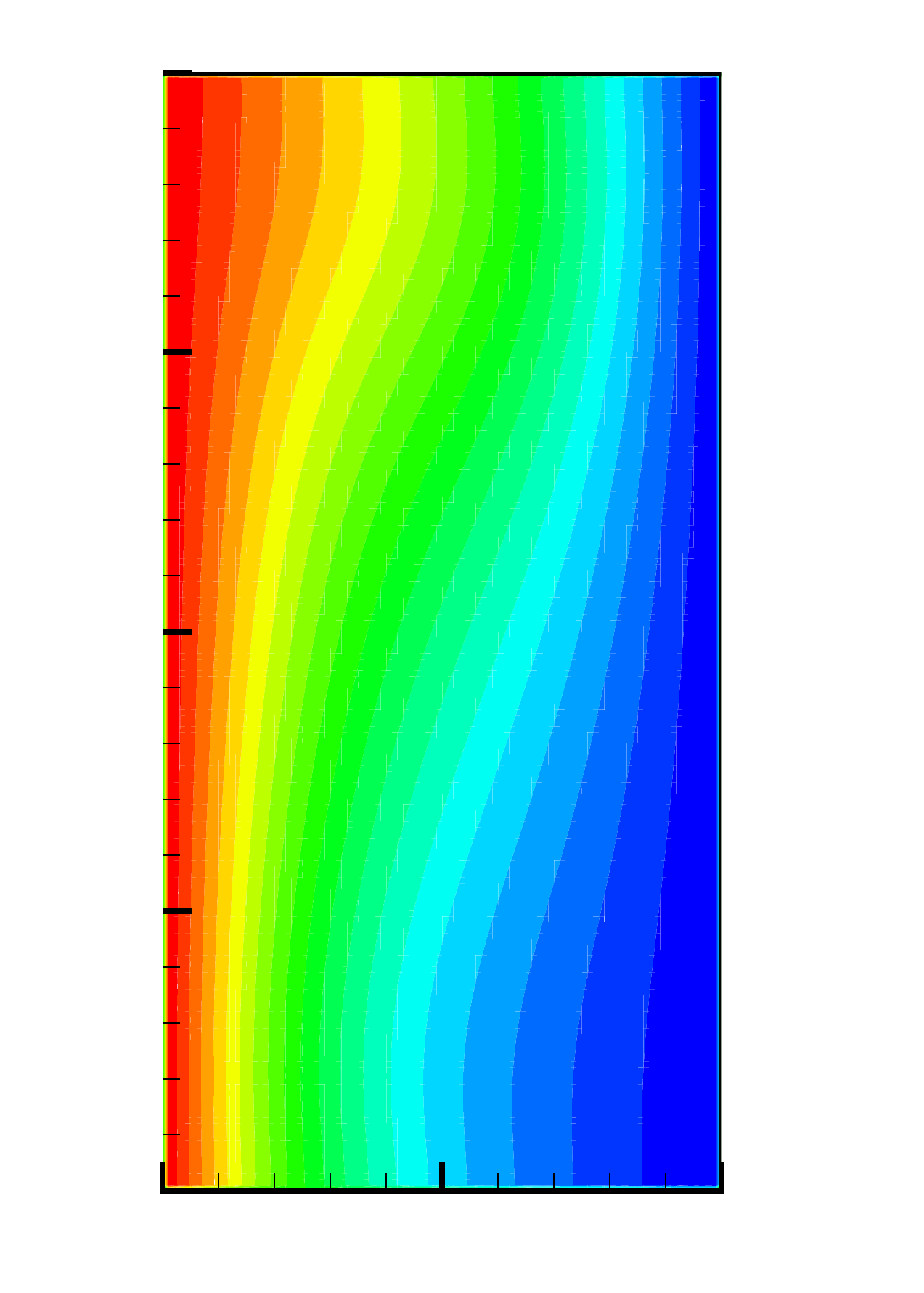}
         }
             \subfloat[Re=$10^4$\label{subfig-2:dummy}]{
       \includegraphics[trim={3cm 1cm 7cm 1.3cm},clip,scale=0.3] {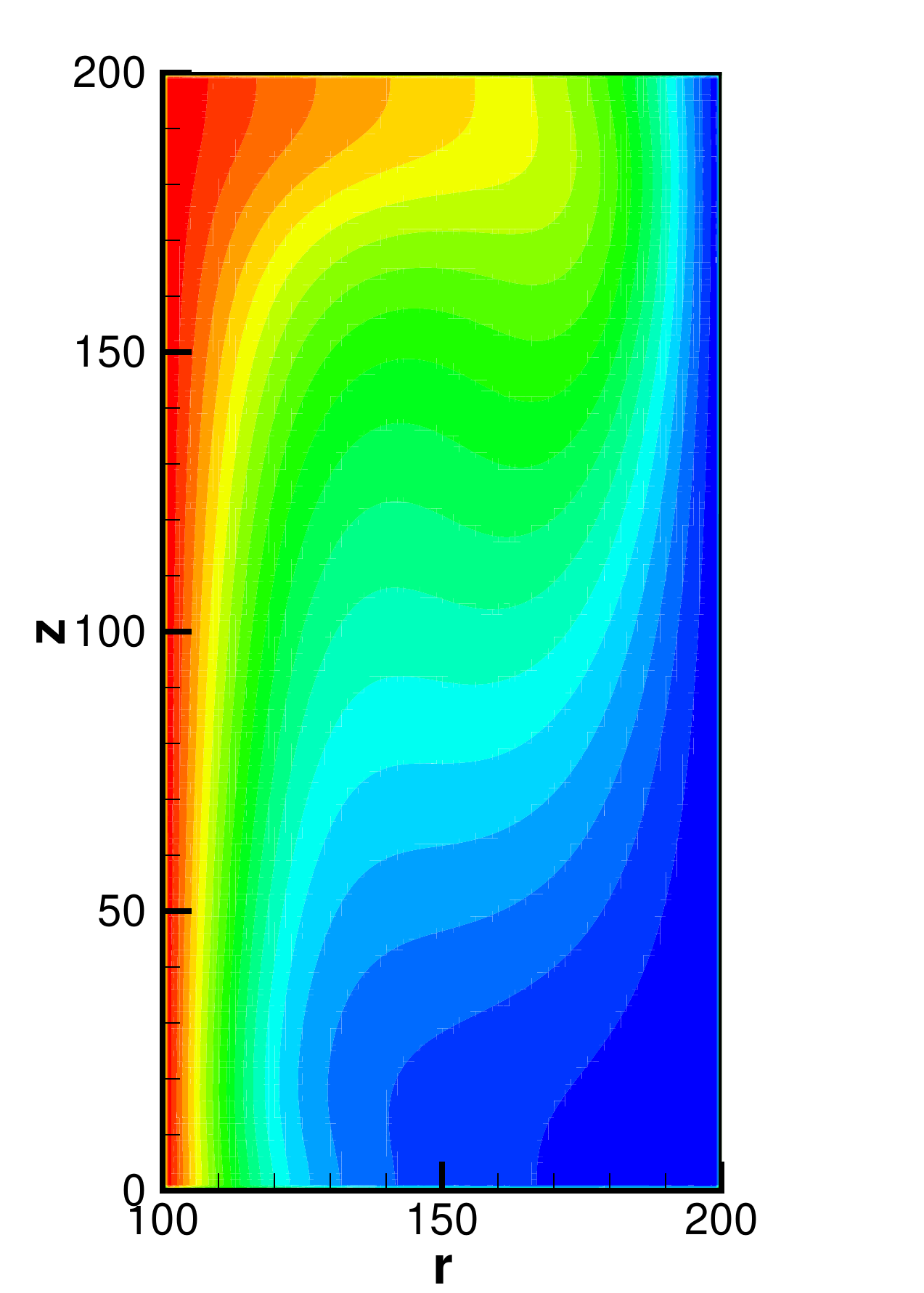}
        }
    \subfloat[Re=$10^5$\label{subfig-2:dummy}]{
        \includegraphics[trim={3cm 1cm 7cm 1.3cm},clip,scale=0.3] {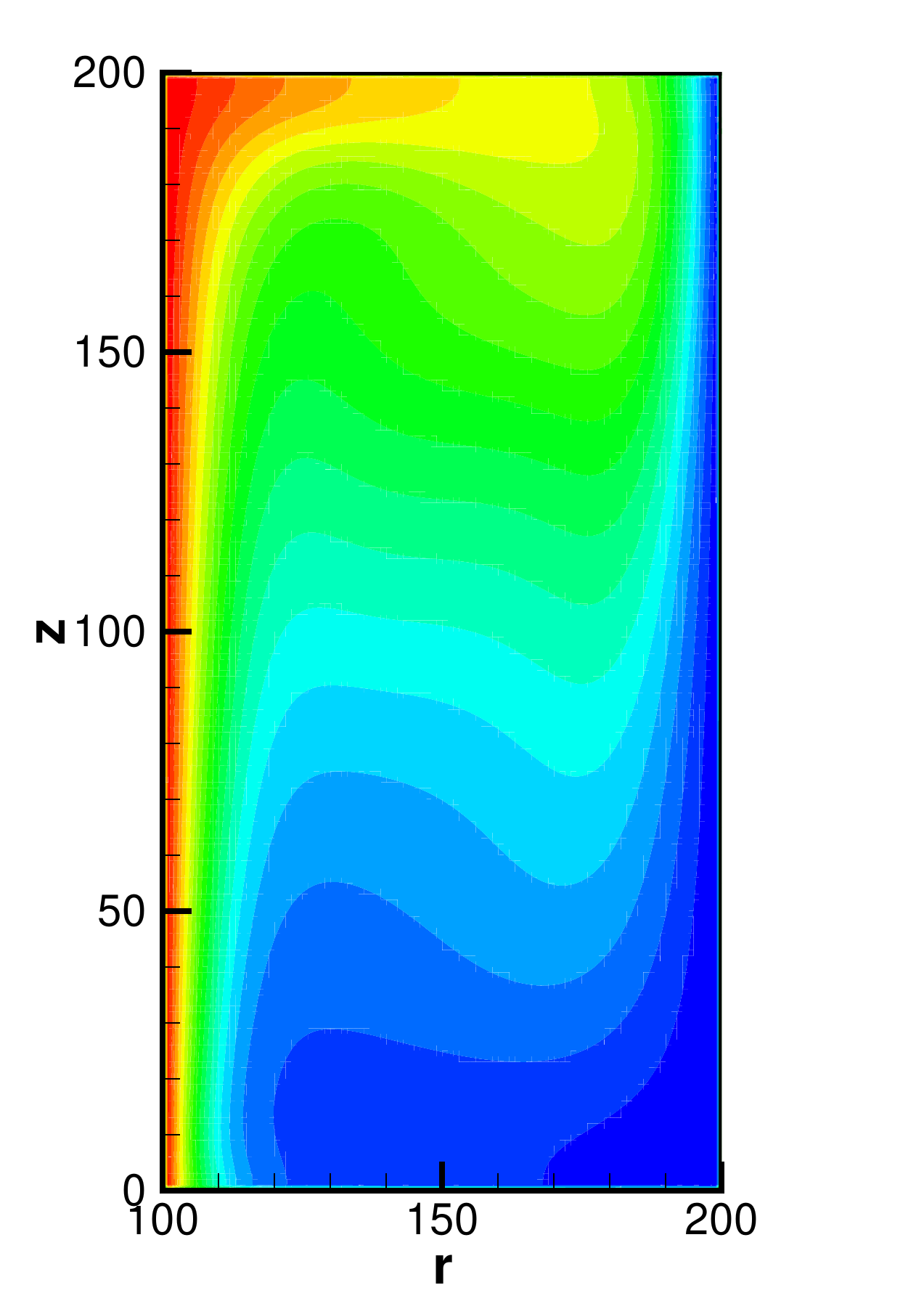}
         }
        \caption{streamlines and isotherms for the natural convection between two co-axial vertical cylinders at $Pr=0.7$ and (a,d) $Ra=10^3,$  (b,e) $Ra=10^4$ and (c,f) $Ra=10^5$ computed using cascaded LB schemes. Top row presents streamlines and the bottom row the isotherms.}
    \label{fig:Natural1}
\end{figure}

It is clear that as $Ra$ increases, the vortical patterns turn to be progressively more complex, with the $Ra=10^5$ case generating additional pairs of vortices around the middle of the annulus. Furthermore, as $Ra$ increase, the isotherms are greatly distorted, and the velocity and thermal boundary layers become thinner near the hot and cold lateral walls signifying the strengthened convection mode of heat transfer. It may be noted that all these observations are consistent with prior studies based on other numerical methods (e.g.,~\cite{Kumar1991,Venkatachalappa2001,li2013multiple}). Then in order to quantify the rates of heat transfer on the lateral walls, the overall Nusselt numbers $Nu_i$ and $Nu_o$ on the inner and outer cylinders can be defined as
\begin{equation*}
Nu_i=\frac{-R_i}{H(T_H-T_L)}\int_o^H{(\partial_y T_i)dx},\, Nu_o=\frac{-R_o}{H(T_H-T_L)}\int_o^H{(\partial_y T_o)dx},
\end{equation*}
 and hence the average Nusselt number $\overline{Nu}=(Nu_i+Nu_o)/2$. Table 1 shows a comparison of the average Nusselt number computed using the cascaded LB scheme for $Ra=10^3, 10^4, and 10^5$ against prior numerical benchmark results~\cite{Kumar1991,Venkatachalappa2001,li2013multiple}. It can be seen that our predictions for the average Nusselt numbers agree well with those obtained by other methods.

\begin{table}[htbp]
\centering
\caption{Comparison of the average Nusselt number $\overline{Nu}$ for different $Ra$ for natural convection in a cylindrical annulus computed using axisymmetric cascaded LB schemes with other reference numerical solutions.}
\label{table:Natural}
\begin{tabular}{c c c c c }
\hline\hline
$Ra$   & Cascaded LB schemes & Ref.~\cite{Kumar1991}& Ref.~\cite{Venkatachalappa2001}&Ref.~\cite{li2013multiple}\\ \hline
$10^3$   &1.688        &     -      &   -&1.692\\
$10^4$  &3.211         &     3.037      & 3.163 &3.215\\
$10^5$  & 5.781         &     5.760     &  5.882 &5.798\\  \hline\hline
\end{tabular}
\end{table}

\subsection{Rayleigh-Benard convection in a circular vertical cylinder}
We now demonstrate the ability of our axisymmetric cascaded LB schemes to simulate Rayleigh-Benard convective in a vertical cylinder, which is classical thermally-driven flow and has been well studied experimentally and using conventional numerical methods (e.g. \cite{liang1969buoyancy,lemembre1998laminar}). Here, the fluid is heated from below, where the bottom wall is at temperature $T_H$ while the top wall is kept at a lower temperature $T_L$ and lateral wall of a cylinder of radius $R$ and height $H$ is maintained to be adiabatic (see Fig.~4). As a result of the buoyancy force generated, this sets up natural convection currents, whose dynamics is governed by the Rayleigh number $Ra=g\beta (T_H-T_L)H^3/(\alpha \nu)$, Prandtl number, $Pr=\nu/\alpha$ and the cylinder aspect ratio $R_A=H/R$. For the purpose of validating the novel LB schemes presented in this work, we set $Pr=0.7, Ra=5\times 10^3$ and $R_A=1$ and the domain is resolved by $100\times 100$ lattice nodes by using the relaxation times $\omega_j=1/\tau, j=4,5$, where $\tau=0.85$. No slip, and constant temperature and adiabatic boundary conditions are represented by using the same approaches are mentioned earlier, and the axis of symmetry is taken into account by using the mirror boundary conditions for the particle distribution functions in the LB schemes.
\begin{figure}[htbp]
\centering
    \subfloat{
        \includegraphics[scale=0.5] {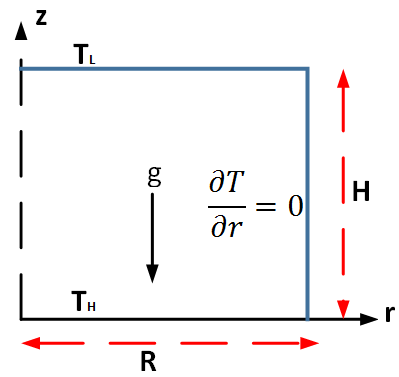}
         }
                    \caption{Schematic illustration of Rayleigh-Benard Convection in a vertical cylinder.}
                    \label{fig:Benard}
    \end{figure}

Figure 5 show the steady isotherms and velocity vectors. Interestingly, it can be seen that based on the initial conditions for temperature, different flow patterns and isotherms are observed. In particular, if the initial temperature is set to $T_L$ everywhere, an up flow draft around the center of the cylinder is observed, while if a higher buoyancy force is prescribed by the initial conditions, a down flow convective current around the center of the cylinder is set up. Those reversal in flow patterns are consistent with findings based on other numerical schemes \cite{zheng2010lattice,lemembre1998laminar}.

\begin{figure}[htbp]
\centering
\advance\leftskip-.5cm
    \subfloat[Up-flow\label{subfig-2:dummy}]{
        \includegraphics[scale=0.25] {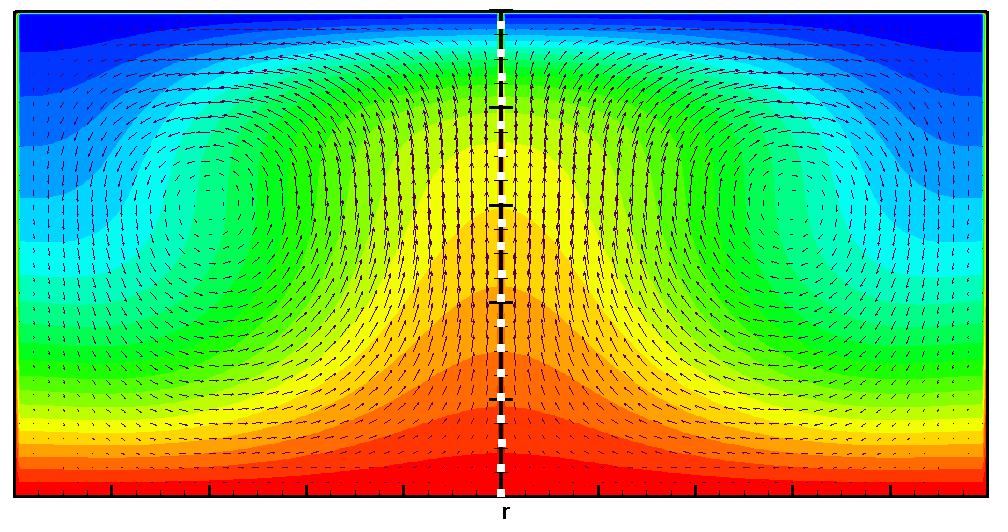}
         }
             \subfloat[Down-flow\label{subfig-2:dummy}]{
        \includegraphics[scale=0.25]  {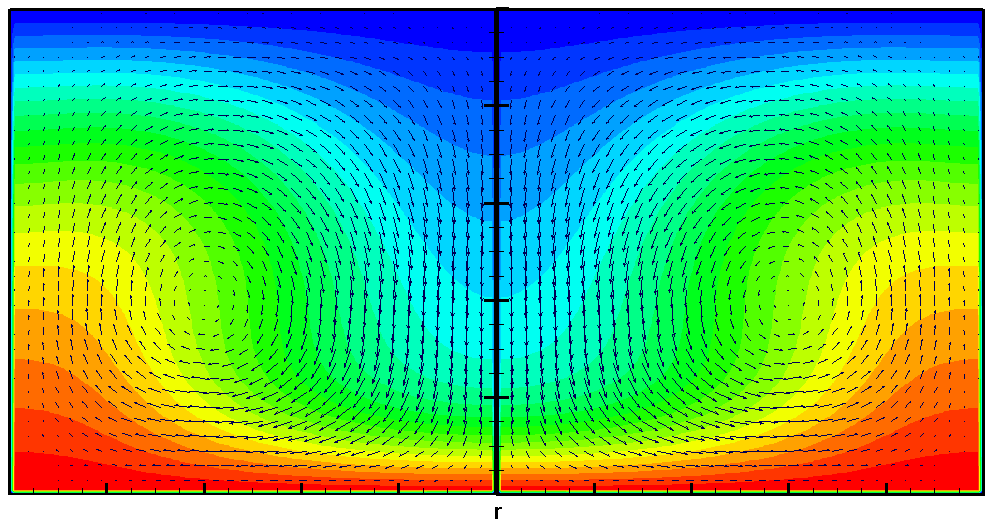}
        }
        \caption{Isotherms  and velocity vectors  for Rayleigh-Bernard convection in a vertical cylinder at $Ra=5\times 10^3$, $Pr=0.7$, $R_A=1$. Left column:~Up-flow pattern and Right column:~Down-flow pattern.}
         \label{fig:Benard1}
\end{figure}

In addition, Table 2 presents a comparison of the maximum velocity in dimensionless form using the natural convection velocity scale $\sqrt{g\beta H (T_H-T_L)}$ obtained using axisymmetric cascaded LB approach with results from the work of \cite{zheng2010lattice,lemembre1998laminar}. Evidently, the computed results agree very well with those reported in the literature.
\begin{table}[]
\centering
\caption{Comparison of the dimensionless maximum velocity obtained using the scale $\sqrt{g\beta H (T_H-T_L)}$ for Rayleigh Benard convection in a vertical cylinder at $Ra=5\times 10^3$ computed using axisymmetric cascaded LB schemes with reference data.}
 \label{table:Benard}
\begin{tabular}{c c c c}
\hline\hline
       $Ra$         & Reference  & Up-flow & Down-flow \\ \hline
    & Ref.~\cite{zheng2010lattice}    & 0.353   & 0.351     \\
        $5\times 10^3$              &  Ref.~\cite{lemembre1998laminar} & 0.353   & 0.353     \\
                      & Present Work  & 0.353   & 0.351     \\ \hline
\end{tabular}
\end{table}

\subsection{Swirling flow in a lid-driven cylindrical container}
In this section, we investigate the ability of the axisymmetric cascaded LB schemes to accurately simulate the dominant role played by the swirling motion and its coupling with the complex radial and axial flow induced in the meridian plane. In this regard, we consider the symmetry breaking flow in a cylindrical container of radius $R$ and height $H$ driven by a rotating top end wall at angular velocity $\Omega$ (see Fig.~6).
\begin{figure}[htbp]
\centering
    \subfloat{
        \includegraphics[scale=0.4] {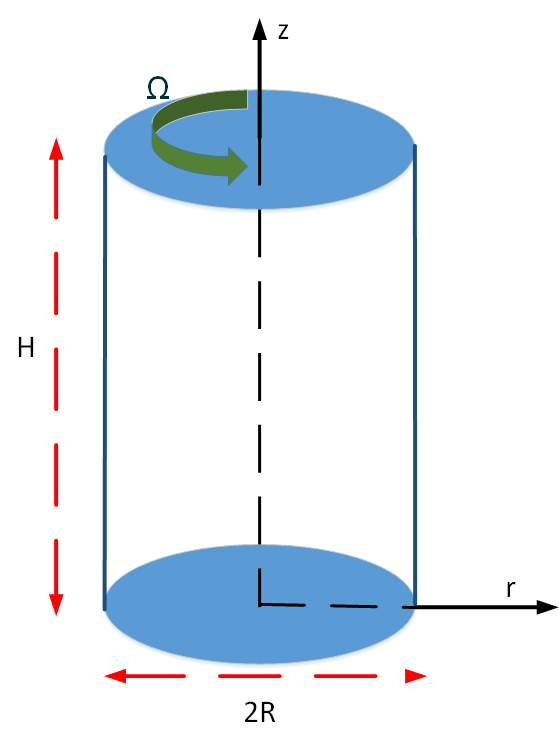}
         }
                    \caption{Schematic of swirling flow in a confined cylinder driven by a rotating top lid.}
                     \label{fig:Cylindrical}
    \end{figure}

 The dynamics of this flow is presented by Eqs. (3a)-(3c), (4a)-(4c), (6) and (7), whose solution scheme via our cascaded LB formulation is presented in Sec.2.2 and 2.3. Briefly, as the fluid in the vicinity of the top lid gains azimuthal motion, it is ejected radially outward, and then downward due to the constraining effect of the side wall. Subsequently as the fluid reaches the bottom it is pushed radially inward, and when it is closer to the axis, it travels upward, thereby completing flow circulation in the meridian plane. The details of the physics and the flow pattern depend on the aspect ratio $R_A=H/R$ and the rotational Reynolds number $Re=R^2 \Omega/\nu$. Various experiments (e.g.,~\cite{hourigan1995spiral, fujimura2001velocity}) and numerical simulations (e.g.,~\cite{gelfgat1996stability,serre2002vortex,blackburn2000symmetry}) have revealed that for certain combinations of the characteristic parameters $R_A$ and $Re$, distinct recirculation regain around the cylinder axis, designated as the vortex breakdown bubble, may occur. For example, Refs.~\cite{fujimura2001velocity,bhaumik2007lattice} show that for cases ($R_A,Re$) equal to (1.5, 990) and (2.5,1010), no vortex breakdown bubbles occur whereas for (1.5, 1290), they do occur.\par
In order to asses and validate our cascaded LB schemes presented earlier to simulate such complex swirling flow, we consider the following four test cases: $Re=990$ and $Re=1290$ with $R_A=1.5$ and $Re=1010$ and $Re=2020$ with $R_A=2.5$. The computational domain is resolved using a mesh resolution of $100\times 150$ for $R_A=1.5$ and $100\times 250$  for $R_A=2.5$. No-slip boundary conditions are used  at bottom, lateral and top walls: $u_\theta=u_r=u_z=0$ at $z=0$ and $r=R$, and $u_\theta=r\Omega$, $u_r=u_z=0$ at $z=H$. The streamlines computed using the cascaded LB schemes for the above four cases are in Fig.~7. It can be seen that no vortex break-down bubbles appear for ($R_A,Re$) equal to (1.5, 990) and (1.5, 1010). On the other hand, one vortex break down bubble is seen at (1.5, 1290) and two break down bubbles occur in the vicinity of the cylinder axis. These distinct regimes in swirling flows and the complex flow structure for different ($R_A,Re$) cases are strikingly consistent with prior numerical solution (e.g.,~\cite{bhaumik2007lattice,guo2009theory,li2010improved,Zhou2011}).
\begin{figure}[htbp]
\centering
\advance\leftskip-2cm
    \subfloat[$R_A=1.5$,~Re=990\label{subfig-2:dummy}]{
        \includegraphics[scale=0.3] {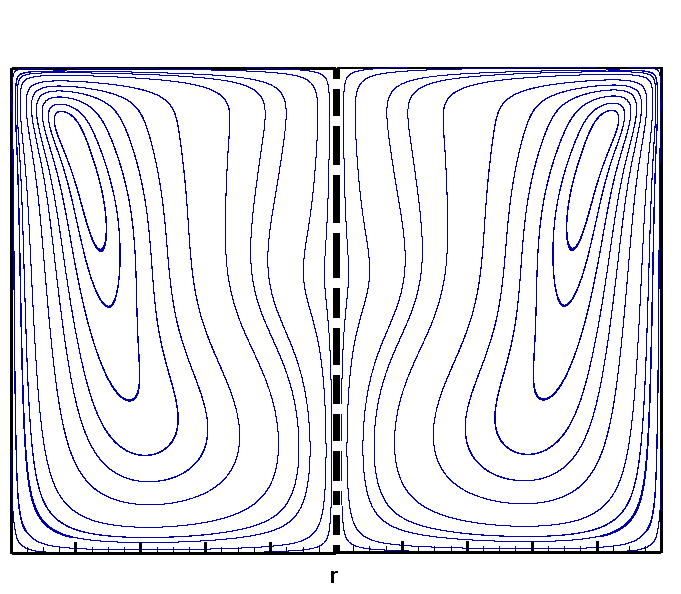}
         }
             \subfloat[$R_A=1.5$,~Re=1290 \label{subfig-2:dummy}]{
       \includegraphics[scale=0.3] {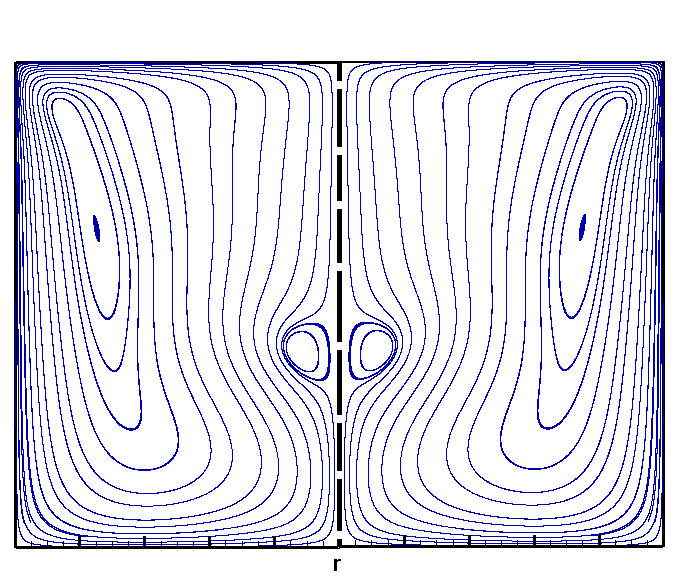}
        }
         \\
          \subfloat[$R_A=2.5$,~Re=1010\label{subfig-2:dummy}]{
        \includegraphics[scale=0.5] {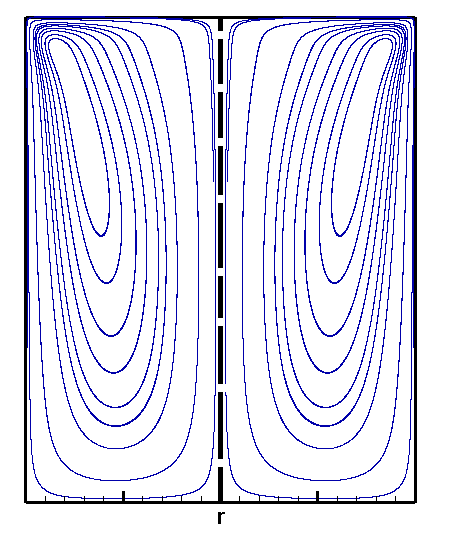}
         }
             \subfloat[$R_A=2.5$,~Re=2200 \label{subfig-2:dummy}]{
       \includegraphics[scale=0.5] {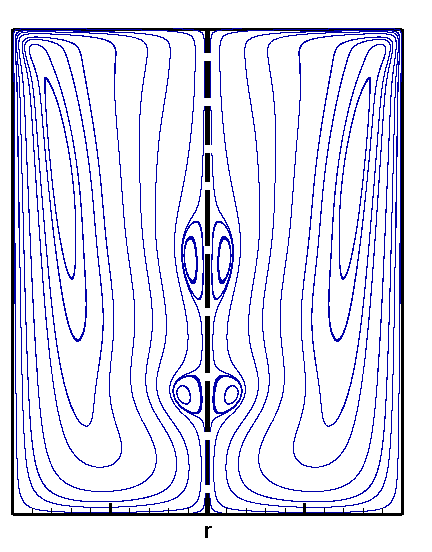}
        }
             \caption{Computed streamline patterns in the meridian plane due to swirling flow in a confined cylinder driven by a rotating lid at various aspect ratios and Reynolds numbers using the axisymmetric cascaded LB sachems: (a) $R_A=1.5$ and $Re=990$, (b) $R_A=1.5$ and $Re=1290$ (c)$R_A=2.5$ and $Re=1010$ and (d)$R_A=2.5$ and $Re=2200$.}
                \label{fig:Cylindrical1}
    \end{figure}
Quantitative comparison of the computed structure of the axial velocities along the axis of symmetry obtained using the axisymmetric cascaded LB schemes for the above four sets of the aspect ratios $R_A$ and Reynolds number $Re$ against the results from a NS-based solver (given in~\cite{bhaumik2007lattice}) are shown in Fig.~8. Here, the axial velocity is scaled by the maximum imposed azimuthal velocity $u_o=\Omega R$ on the rotating lid and the axial distance $z$ by the cylinder height $H$. The numerical results of our central moments based  cascaded LB method for the axial velocity profiles are in very good agrement with the NS-based solution approach~\cite{bhaumik2007lattice}. Also, in particular, notice local negative values for the axial velocities for the cases $Re=1290$ and $R_A=1.5$ and $Re=2200$ and $R_A=2.5$, which is an indication of the presence of one or more vortex breakdown bubbles. As such, both the magnitudes and the shapes of the axial velocity distributions are well reproduced by our cascaded LB approach using operator splitting to represent complex flows in cylindrical coordinates.

\begin{figure}[htbp]
    \subfloat[Re=990, $R_A=1.5$\label{subfig-2:dummy}]{
        \includegraphics[scale=0.45] {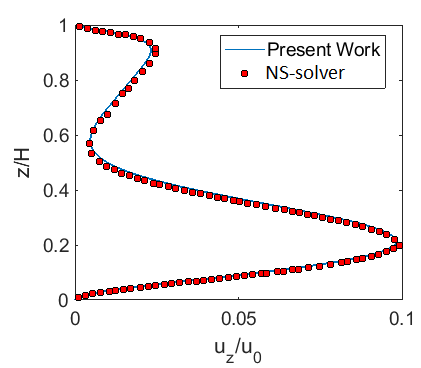}
         }
             \subfloat[Re=1290, $R_A=1.5$\label{subfig-2:dummy}]{
       \includegraphics[scale=0.45] {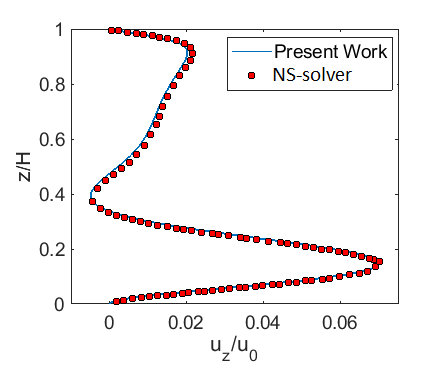}
        }
         \\
          \subfloat[Re=1010, $R_A=2.5$\label{subfig-2:dummy}]{
        \includegraphics[scale=0.45] {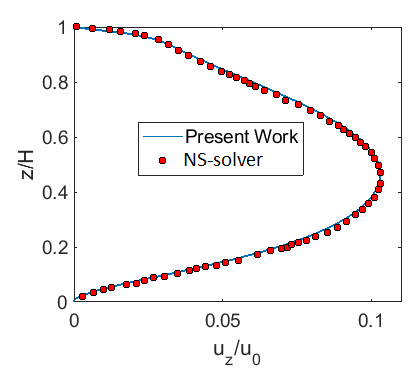}
         }
             \subfloat[Re=2200, $R_A=2.5$\label{subfig-2:dummy}]{
       \includegraphics[scale=0.45] {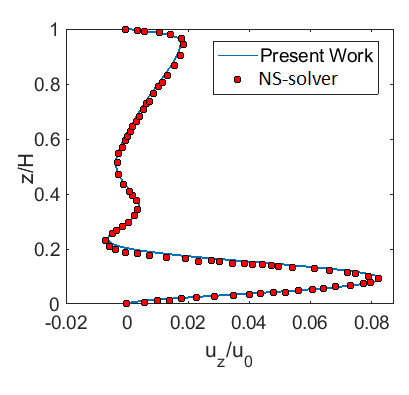}
        }
             \caption{Dimensionless axial velocity profile $u_z/u_o$ as a function of the dimensional axial distance $z/H$ for (a) $R_A=1.5$ and $Re=990$, (b) $R_A=1.5$ and $Re=1290$ (c)$R_A=2.5$ and $Re=1010$ and (d)$R_A=2.5$ and $Re=2200$: Comparison between axisymmetric cascaded LB scheme predictions and NS-based solver results (\cite{bhaumik2007lattice})}
   \label{fig:Cylindrical2}
    \end{figure}

\subsection{Mixed convection in a slender vertical annulus between two coaxial cylinders}
We will now assess our new axisymmetric LB computational approach based on central moments to simulate the combined effects of rotation and buoyancy forces on the flow and heat transfer in confined cylindrical spaces. In this regard, we investigate mixed convection in a slender vertical annulus between two coaxial cylinders arising due to inner side wall rotation, which  has numerous applications related to rotating machinery and various other heat transfer systems. This problem involving both natural convection and forced convection due to rotation can test all the three axisymmetric cascaded LB formulations (Secs.~2.2-2.4) in a unified manner. \par A schematic arrangement of this axisymmetric thermal flow problem is shown in Fig.~9.
\begin{figure}[htbp]
\centering
    \subfloat{
        \includegraphics[scale=0.4] {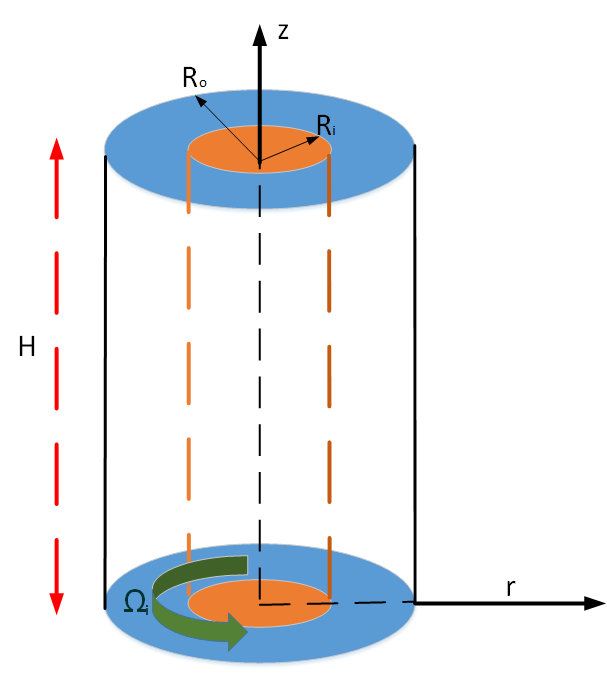}
         }
   \caption{Schematic of the arrangement for mixed convection in a slender cylindrical annulus with inner lateral wall rotation.}
       \label{fig:Mixed}
    \end{figure}
It consist of two coaxial cylinders of height $H$, with an annular gap $D=R_o-R_i$, where $R_i$ and $R_o$ are the radii of the inner and outer cylinders, respectively. The lateral walls of the inner and outer cylinders are maintained at temperatures $T_H$ and $T_L$, respectively, where $T_H>T_L$, and their bottom and top ends are thermally insulated. The inner cylinder is subjected to rotation at an angular velocity $\Omega_i$, while the outer cylinder and the end walls are considered to be rigidly fixed. As noted in a recent study \cite{wang2014fractional}, this problem is governed by the following characteristic dimensionless parameters: Prandtl number $Pr=\nu/\alpha$, radius ratio $R_{io}=R_o/R_i$ slenderness ratio $\eta=H/(R_o-R_i)$, Reynolds number $Re=\Omega_iR_iD/\nu$, Grashof number $Gr=g\beta (T_H-T_L)D^3/\nu^2$, and $\sigma=Gr/Re^2$, where the parameter $\sigma$ is used to measure the strength of the buoyancy force relative to the centrifugal force. Hence, $\sigma$ characterizes the degree of mixed convection.

In the present study, we set $Pr=0.7$, $R_{io}=2$, $\eta=10$, $Re=100$, and three cases of $\sigma$ are considered: $\sigma=0,~0.01$ and $0.05$. The grid resolution used for all the three cases is $40\times400$, in which the location of the inner cylinder from the axis $R_i$ is at 40. Figure 10 shows the computed contours of the azimuthal velocity, temperature field, vorticity and streamlines for the above three values of $\sigma$.
\begin{figure}[htbp]
\centering
    \subfloat{
        \includegraphics[height=6.8cm, width=9cm]{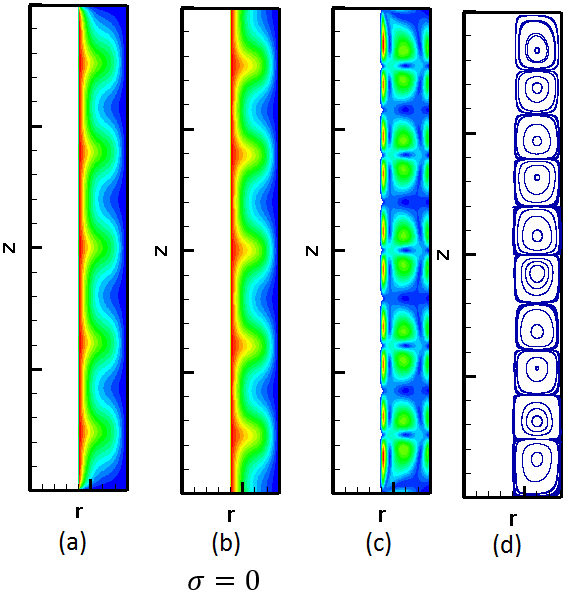}
         }
                               \\
                          \subfloat{
        \includegraphics[height=6.8cm, width=9cm] {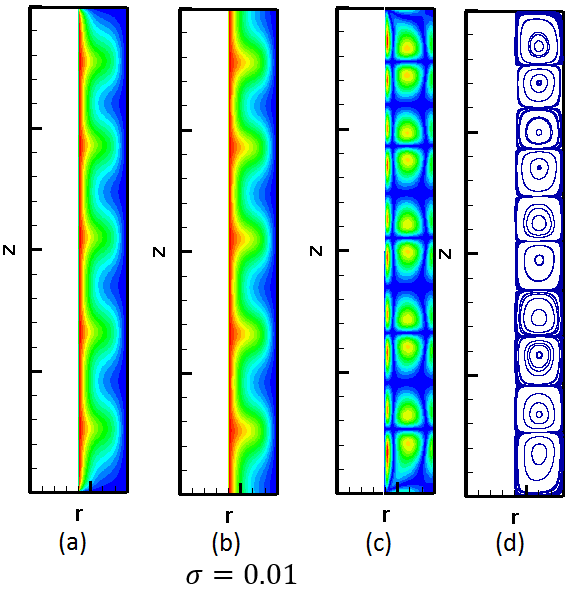}
         }
               \\
                     \subfloat{
        \includegraphics [height=6.8cm, width=9cm] {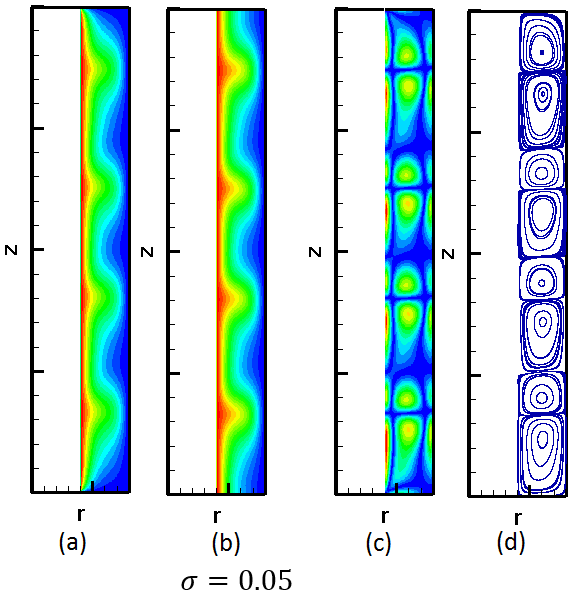}
         }
                              \caption{Contours of (a) azimuthal velocity, (b) temperature, (c) vorticity, and (d) streamlines for mixed convection in a slender cylindrical annulus for three different values of $\sigma$ computed using the axisymmetric cascaded LB schemes.}
    \label{fig:Mixed1}
\end{figure}
When $\sigma=0$, there is no buoyancy force and the flow and the temperature fields are influenced by the centrifugal force and the forced convection effects, which manifest in the form of five pairs of  counter-rotating cells, viz., the classical Taylor vortex cells arising from centrifugal flow instability between curved walls \cite{koschmieder1993benard}. As $\sigma$ is increased, the presence of buoyancy forces and the associated natural convective fluid  currents alter the overall flow structure and the temperature field by their complicated interactions with primary vortex cells induced by the swirling effects from inner wall rotation. For example, when $\sigma=0.05$, a four-pairs based Taylor vortex structure, rather than  five-pair of vortex cells observed for $\sigma=0$, arises from the relative weakening effects of the centrifugal forces in the presence of heating . The strength of the Taylor vortex in the positive azimuthal direction $\theta$ is seen to be enhanced, while that negative $\theta$ direction appear to be diminished and these observations are consistent with the benchmarks results \cite{Ho1993,Ball1987} and recent numerical simulations \cite{wang2016lattice}. In order to quantify the heat transfer rate in the presence of mixed convection, a mean equivalent thermal conductivity at the inner cylinder can be defined as
\begin{eqnarray*}
\overline {k}_{eq}|_i=\frac{ln R_{io}}{\mu}\int_o^H {\left(-r \frac{\partial T}{\partial r}|_{r=R_i}\right)dr}.
\end{eqnarray*}
Table 3 presents a comparison of the equivalent thermal conductively computed using the axisymmetric cascaded LB formulations against the benchmark results \cite{Ho1993,Ball1987} for different values of $\sigma$. Very good quantitative agrement is seen and this validates the ability of the cascaded LB schemes in the cylindrical coordinate system to represent complex flows with heat transfer.
\begin{table}[H]
\centering
\caption{Comparison of the mean equivalent thermal  conductivity at the inner cylinder in a  slender vertical cylindrical annulus during mixed convection for $Re=100, Pr=0.7, R_{io}=2, \eta=10$ at different values of $\sigma$.}
 \label{table:Mixed}
\begin{tabular}{c c c c}
\hline\hline
 $\sigma$       &  Ref.~\cite{Ball1987}  &  Ref.~\cite{Ho1993}& Present Work \\ \hline
 0 & 1.473  & 1.393&   1.395  \\
  0.01&1.370&  1.383 & 1.378    \\
 0.05 &1.324     & 1.323 &  1.321   \\ \hline
\end{tabular}
\end{table}

 \subsection{Melt flow and convection during Czochralski crystal growth in a rotating cylindrical crucible }

As the last test problem, we simulate melt flow and convection during Czochalski crystal growth, based on a configuration reported by Wheeler~\cite{wheeler1990four}, using our axisymmetric cascaded LB schemes. This Wheeler's benchmark problem involved both forced convection due to the rotation of the crucible and the crystal and natural convection arising from heating effects in the presence of gravity. It has been studied by a variety of numerical schemes (e.g.,~\cite{Xu1997,shu1997efficient,peng2003numerical,wang2014fractional}). The geometric arrangement of this problem is shown in Fig.~11.

\begin{figure}[htbp]
\centering
    \subfloat{
        \includegraphics[scale=0.5] {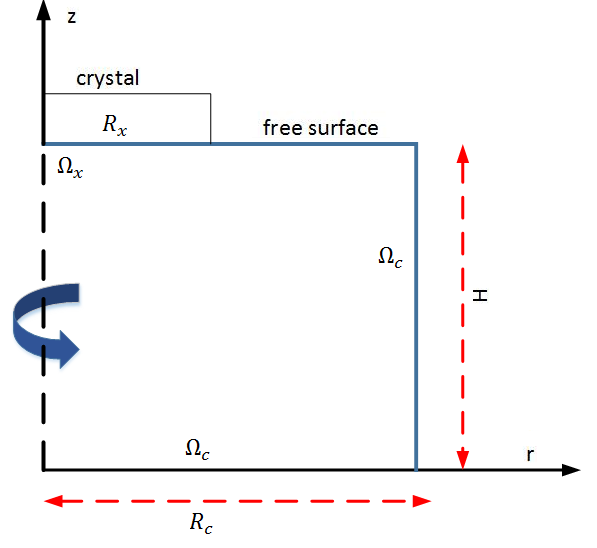}
         }
   \caption{Geometric arrangement of melt flow and convection during Czochralski crustal growth in a rotating crucible$\--$Wheeler's benchmark problem.}
       \label{fig:Wheeler}
    \end{figure}
 Liquid melt in a cylindrical rotating crucible of radius $R_c$ and height $H$ at an angular rotation rate of $\Omega_c$ undergoes stirred vortical motion in the meridian plane, which is aided by the angular rotation of the solid crystal of radius $R_x$ at rate $\Omega_x$. In addition, natural convection is set up due to the buoyancy force generated from a differential heating, where the bottom is insulated and its crucible side is maintained at a temperature $T_H$, while the crystal is at a lower temperature $T_L~(i.e.,~T_L<T_H)$. These can be prescribed in terms of the following boundary conditions, where the ($x,z$) coordinates are scaled by $R_c:$

 \begin{subequations}
\begin{flalign*}
 &u_r=u_\theta=\dfrac{\partial u_z}{\partial r}=\dfrac{\partial T}{\partial r}=0 \; \;\; &\text{for} \; \; \; r=0 \;\; 0\leq z \leq \alpha\\
 &u_r=u_z= 0,\; u_\theta=\Omega_cR_c,\; T=T_H\; \;\; &\text{for} \; \; \; r=1 \;\; 0\leq z \leq \alpha\\
 &u_r=u_z= 0,\; u_\theta=r\Omega_c, \;\dfrac{\partial T}{\partial z}=0\; \;\; &\text{for} \; \; \; z=0 \;\; 0\leq r \leq 1\\
 &u_r=u_z=0, \; u_\theta=r\Omega_x,\; T=T_L \; \;\; &\text{for} \; \; \; z=\alpha \;\;0\leq r \leq \beta\\
 &\dfrac{\partial u_r}{\partial r}= \dfrac{\partial u_\theta}{\partial z}=0,\; u_z=0, \; T=T_L+\dfrac{r-\beta}{1-\beta}(T_H-T_L)\; \;\; &\text{for} \; \; \; z=\alpha\;\;  \beta\leq r \leq 1
\end{flalign*}
\end{subequations}
 where $\alpha=H/R_c, \beta=R_x/R_c$. This flow problem is characterized by the following dimensionless parameters: Reynolds numbers due to crucible and crystal rotations $Re_c=R_c^2\Omega_c/\nu$ and $Re_x=R_x^2\Omega_x/\nu$, and Prandtl number $Pr=\nu/\alpha$. We investigate the ability of the axisymmetric cascaded LB schemes for the simulation of mixed convection associated with the Wheeler's benchmark problem for the following two cases: (a) $Re_x=100$, $Re_c=-25$ and (b) $Re_x=1000$, $Re_c=-250$, where the negative sign denotes that the sense of rotation of the crystal is apposite to that of the crucible. We take $Pr=0.05,~\alpha=1$, and $\beta=1$ and use a grid resolution of $100\times 200$ for the simulation of both the cases.\par
Figure 12 shows the streamlines and isotherm contours in the meridian plane of the liquid melt motion for the two cases. It can be seen that a recirculating vortex appears around the upper left region below the crystal in both cases in addition to the primary vortex. The center of this secondary vortex is found to move to the right at higher Reynolds numbers as a result of higher associated centrifugal forces. On the other hand, the forced convection has modest effect on the temperature distribution, as they are largely alike for both the cases due to the relatively low Reynolds numbers considered.
\begin{figure}[htbp]
\centering
\advance\leftskip-2cm
    \subfloat[$Re_x=100 , Re_c=-25$\label{subfig-2:dummy}]{
        \includegraphics[scale=0.4] {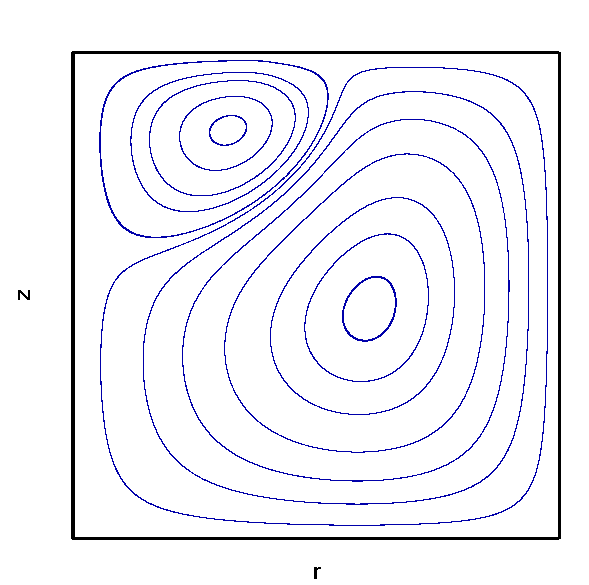}
         }
             \subfloat[$Re_x=1000 , Re_c=-250$\label{subfig-2:dummy}]{
       \includegraphics[scale=0.4] {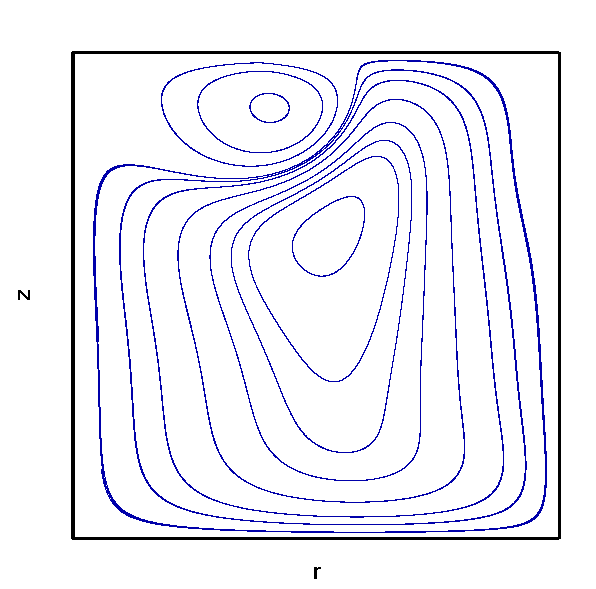}
        }
         \\
          \subfloat[$Re_x=100 , Re_c=-25$\label{subfig-2:dummy}]{
        \includegraphics[scale=0.4] {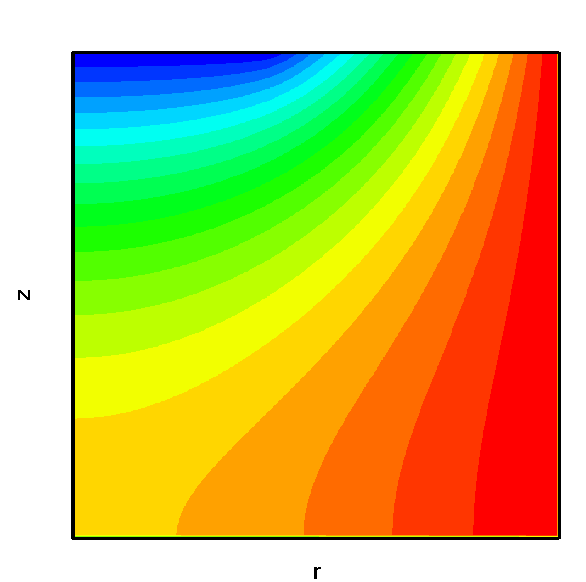}
         }
             \subfloat[$Re_x=1000 , Re_c=-250$\label{subfig-2:dummy}]{
       \includegraphics[scale=0.4] {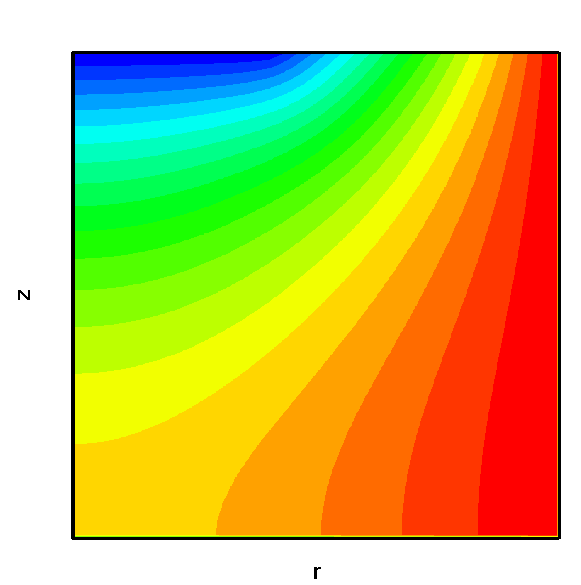}
        }
             \caption{Streamlines (upper row) and isotherms (bottom row) corresponding to two cases of the Wheeler's benchmark problem of melt flow and convection during Czochralshi crystal growth: $Re_x=100$, $Re_c=-25$ (left) and $Re_x=1000, Re_c=-250$ (right). }
              \label{fig:Wheeler1}
    \end{figure}

Table 4 shows the computed absolute maximum values of the streamfunction $\psi_{max}$ for the above two cases and compared with prior numerical results presented in \cite{peng2003numerical,Xu1997}. In the pseudo-2D Cartesian coordinates, this is obtained by solving for $\psi$ using $\partial \psi/\partial y=-yu_x$ and $\partial \psi/\partial x=yu_y$. The good agreement confirms that the new axisymmetric cascaded LB schemas presented in this study can effectively simulate complex flow and heat transfer problems in cylindrical geometries.

\begin{table}[htbp]
\centering
\caption{Comparison of the maximum value of the stream function $\psi_{max}$ computed using the axisymmetric cascaded LB schemes with reference numerical solutions for the Wheeler's benchmark problem.}
\label{Wheeler}
\begin{tabular}{c c c}
\hline\hline
{Reference}    & {$Re_x=10^2$,   $Re_c$=-25} & {$Re_x=10^3$,  $Re_c$=-250} \\ \hline
{Present Work} & {0.1183}                                 &1.123 \\
Ref.~\cite{Peng2003}                           & 0.1140                                                      & 1.114                                                       \\
Ref.~\cite{Xu1997}                              & 0.1177                                                      & 1.148                                                       \\ \hline
\end{tabular}
\end{table}

\newpage
\section{Summary and Conclusions}
Thermally stratified fluid convection including rotational effects within cylindrical confined spaces represents an important class of flows with numerous engineering applications. Exploiting axial symmetry in such problems leads to their representation in terms of a quasi-2D system of equations with geometric source terms in the meridian plane, which can significantly reduce computational and memory costs when compared to their 3D modeling.\par In this work, we have presented axisymmetric cascaded LB schemes for convecttive flows with combined rotation and thermal stratification effects in cylindrical geometries. A triple distribution functions  based approach is employed in this regard, in which the axial and radial momentum as well as the pressure field are solved using a D2Q9 lattice based cascaded LB scheme, while the azimuthal momentum and the temperature field are solved using the two other cascaded LB schemes, each based on a D2Q5 lattice. The collision step in these three schemes is based on the relaxation of different central moments at different rates to represent the dynamics of the fluid motion as well as the advection-diffusion transport of the passive scalar fields in a consistent framework. The geometric mass, momentum and energy source terms arising in the quasi-2D formulation are incorporated using a simpler operator splitting based approach involving a symmetric application of their effects given in terms of appropriate change of moments for two half time steps around the collision step. This new computational approach is then used to simulate a variety of complex axisymmetric benchmark thermal flow problems including natural convection between two coaxial cylinders, Rayleigh-Benard convection in a vertical cylinder, mixed convection in a slender vertical annulus between two cylinders under combined rotation and buoyancy forces, and convective flow of a melt during Czochralski crystal growth in a rotating cylindrical crucible. Comparison of the computed results obtained using the axisymmetric cascaded LB schemes for such thermal convective flows for the structures of the flow and thermal fields, as well as the heat transfer rates given in terms of the Nusselt number against prior benchmark numerical solutions demonstrate their good accuracy and validity.

\end{document}